\documentclass{article}

\usepackage[dvipdfmx]{graphicx}
\usepackage{amsmath}        
\usepackage{amssymb}        
\usepackage{authblk}      
\usepackage{cite}
\usepackage[dvips]{color}
\usepackage{fancyhdr}

\def\erase#1{{}}
\def\EqArrerase#1{{}}

\makeatletter
 \renewcommand{\theequation}{%
 \thesection.\arabic{equation}}
 \@addtoreset{equation}{section}
\makeatother

\def\T{{\rm T}}

\def\GL{{G\kern-.12em L\kern-.04em}}
\def\OSp{{O\kern-.11em S\kern-.04em p}}
\def\IOSp{{I\kern-.06em O\kern-.11em S\kern-.04em p}}
\def\MN{{M\kern-.14em N}}
\def\NM{{N\kern-.14em M}}
\def\NL{{N\kern-.14em L}}
\def\LN{{L\kern-.11em N}}
\def\ML{{M\kern-.14em L}}
\def\LM{{L\kern-.11em M}}
\def\RN{{R\kern-.11em N}}
\def\NR{{N\kern-.14em R}}
\def\RM{{R\kern-.11em M}}
\def\MR{{M\kern-.14em R}}
\def\RL{{R\kern-.11em L}}
\def\LR{{L\kern-.11em R}}
\def\RS{{R\kern-.11em S}}
\def\SR{{S\kern-.11em R}}
\def\SN{{S\kern-.11em N}}
\def\NS{{N\kern-.11em S}}
\def\SM{{S\kern-.11em M}}
\def\MS{{M\kern-.11em S}}
\def\SL{{S\kern-.11em L}}
\def\LS{{L\kern-.11em S}}
\def\sqr#1#2{{\vcenter{\hrule height.#2pt
      \hbox{\vrule width.#2pt height#1pt \kern#1pt
          \vrule width.#2pt}
      \hrule height.#2pt}}}
\def\bra0{\langle0|}
\def\ket0{|0\rangle}
\def\soeji#1_#2#3{#1_{#2}\cdots#1_{#3}}
\def\longgLRarrow{\longleftarrow\kern-3pt\relbar\kern-3pt\relbar\kern-3pt%
\longrightarrow}
\def\longLRarrow{\longleftarrow\kern-3pt\relbar\kern-3pt\longrightarrow}
\def\longLarrow{\longleftarrow\kern-3pt\relbar\kern-3pt\relbar\kern-3pt\relbar}
\def\longRarrow{\relbar\kern-3pt\relbar\kern-3pt\relbar\kern-3pt\longrightarrow}
\def\bothDer#1#2#3{%
\overset{\kern-.7em\stackrel{#1}{#2}}{\partial_{#3}}}
 
\makeatletter
 \renewcommand{\theequation}{%
 \thesection.\arabic{equation}}
 \@addtoreset{equation}{section}
\makeatother

\begin{document}
\thispagestyle{fancy}

\title{Quantum Conformal Gravity}

\author[1]{Ichiro Oda\thanks{ioda@sci.u-ryukyu.ac.jp}}
\author[2]{Misaki Ohta\thanks{350455@uwr.edu.pl}}
\affil[1]{Department of Physics, Faculty of Science, University of the Ryukyus, Nishihara, Okinawa 903-0213, Japan}
\affil[2]{Department of Physics and Astronomy, University of Wroclaw, plac Maksa Borna 9, PL-50204 Wroclaw, Poland}

\date{}

\maketitle

\thispagestyle{fancy}

\begin{abstract}

We present the manifestly covariant canonical operator formalism of a Weyl invariant (or equivalently, 
a locally scale invariant)  gravity whose classical action consists of the well-known conformal gravity 
and Weyl invariant scalar-tensor gravity, on the basis of the Becchi-Rouet-Stora-Tyupin (BRST) formalism.  
It is shown that there exists a Poincar${\rm{\acute{e}}}$-like $\IOSp(8|8)$ global symmetry as in Einstein's general relativity, 
which should be contrasted to the case of only the Weyl invariant scalar-tensor gravity where we have
a more extended Poincar${\rm{\acute{e}}}$-like $\IOSp(10|10)$ global symmetry. This reduction of the global
symmetry is attributed to the presence of the St\"{u}ckelberg symmetry.

\end{abstract}

\section{Introduction}

General relativity (GR) by Einstein is a mathematically beautiful and physically successful classical theory. It is constructed
by two fundamental principles, general coordinate invariance and equivalence principle, and is described in terms
of Riemann geometry where the metric tensor is regarded as fundamental dynamical variables. Since general relativity can 
account for many astrophysical and cosmological phenomena without any conflict with observations and experiments done so far, 
there is no need for modifying it at least at large distance scales.

On the other hand, it is a well-established fact that the physics must be described by quantum field theory (QFT).   
Unfortunately, it seems to be difficult to construct a quantum field theory of general relativity owing to its
nonrenormalizability although the perturbative non-renormalizability has nothing to do with the consistency 
of the theory. In the interest of renormalizability, it is natural to alter the Einstein-Hilbert Lagrangian
by adding to it the most general quadratic Lagrangian ${\cal L}$ of dimension four at most:
\begin{eqnarray}
\frac{1}{\sqrt{-g}} {\cal L} = \frac{1}{16 \pi G} ( R - 2 \Lambda) + \alpha_r R^2 
- \alpha_c C_{\mu\nu\rho\sigma} C^{\mu\nu\rho\sigma},
\label{Ultra-com}  
\end{eqnarray}
which is known as a renormalizable gravitational theory  \cite{Stelle}. However, a notorious problem happens and it is associated 
with the last term involving conformal tensor $C_{\mu\nu\rho\sigma}$: As far as this term exists in the Lagrangian, we have 
a spin-2 massive ghost which makes not only the classical theory be unstable because of unbounded energy from below 
but also the quantum theory be non-unitary owing to the ghost with negative norm.  

At extremely high energies, it is expected that the kinetic term dominates the mass term and as a result
all particles can be effectively regarded as massless particles.  In such a situation, a global or local scale symmetry naturally 
appears in addition to general coordinate invariance. Since the global scale symmetry could be broken by the no-hair theorem
of black holes in a curved space-time \cite{MTW}, it is plausible to suppose that the local scale symmetry, which we call ${\it Weyl \, symmetry}$, 
plays a role at high energies. If we impose the Weyl symmetry on the Lagrangian (\ref{Ultra-com}) and assume that 
Einstein's general relativity is restored at low energies, we would have the following Lagrangian:  
\begin{eqnarray}
\frac{1}{\sqrt{-g}} {\cal L} = \frac{1}{12} \phi^2 R + \frac{1}{2} g^{\mu\nu} \partial_\mu \phi \partial_\nu \phi  
- \alpha_c C_{\mu\nu\rho\sigma} C^{\mu\nu\rho\sigma},
\label{W-Ultra-com}  
\end{eqnarray}
where in the {\it{unitary gauge}} $\phi = \sqrt{\frac{3}{4 \pi G}}$ the terms except for conformal gravity on the
right-hand side (RHS) produce the Einstein-Hilbert term.

In this article, as the first step for understanding the problem of the massive ghost, we wish to construct 
the manifestly covariant canonical operator formalism of the Weyl invariant gravity as defined 
in the Lagrangian (\ref{W-Ultra-com}) on the basis of the Becchi-Rouet-Stora-Tyupin (BRST) formalism \cite{Kugo-Ojima}. 
We will see that this construction is very subtle since we have to carefully pick up gauge fixing conditions 
in order to introduce as many independent BRST transformations as possible.
    
The paper is organized as follows. In Section 2, we review the classical theory where the Lagrangian is
constructed out of the well-known conformal gravity and the Weyl invariant scalar-tensor gravity.
In Section 3, we shed light on quantum aspects of our theory. We will find that the existence of the
St\"{u}ckelberg symmetry makes it impossible to construct three independent BRST transformations
and it allows us to make only two independent BRST transformations. In Section 4, we perform the canonical
quantization. In Section 5, on the basis of the canonical formalism in Section 4, we derive various
equal-time (anti)commutation relations. In Section 6, we analyze asymptotic fields by expanding not only the
metric around a flat Minkowski metric but also the scalar field around a constant background. 
In Section 7, we derive the four-dimensional (anti)commutation relations from the equal-time (anti)commutation
relations and clarify that the physical modes of our theory are composed of a massive ghost with indefinite norm
and a massless graviton, and the other modes belong to the BRST quartets which appear in the physical
subspace only as zero norm states. The final section is devoted to discussion. 

Two appendices are put for technical details. In Appendix A, a derivation of the equal-time commutation
relation between $\dot A_\mu$ and $b_\mu$ is given, and in Appendix B we present various equal-time
(anti)commutation relations which are necessary in deriving the four-dimensional (anti)commutation relations
in Section 7.

\section{Classical theory}

In this section, we consider a classical gravitational theory which is invariant under both general coordinate 
transformation (GCT) and Weyl transformation (or equivalently, a local scale transformation) in four 
dimensional Riemann geometry. Our classical Lagrangian consists of Weyl invariant scalar-tensor gravity \cite{Fujii} and 
conformal gravity\footnote{We follow the notation and conventions of Misner-Thorne-Wheeler (MTW) 
textbook \cite{MTW}. Lowercase Greek letters $\mu, \nu, \dots$ and Latin ones $i, j, \dots$ are used 
for spacetime and spatial indices, respectively; for instance, $\mu= 0, 1, 2, 3$ and $i = 1, 2, 3$. 
The Riemann curvature tensor and the Ricci tensor are, respectively, defined by 
$R^\rho{}_{\sigma\mu\nu} = \partial_\mu \Gamma^\rho_{\sigma\nu} 
- \partial_\nu \Gamma^\rho_{\sigma\mu} + \Gamma^\rho_{\lambda\mu} \Gamma^\lambda_{\sigma\nu} 
- \Gamma^\rho_{\lambda\nu} \Gamma^\lambda_{\sigma\mu}$ and $R_{\mu\nu} = R^\rho{}_{\mu\rho\nu}$. 
The Minkowski metric tensor is denoted by $\eta_{\mu\nu}$; $\eta_{00} = - \eta_{11} = - \eta_{22} 
= - \eta_{33} = -1$ and $\eta_{\mu\nu} = 0$ for $\mu \neq \nu$.} 
\begin{eqnarray}
{\cal L}_0 = {\cal L}_{WIST} + {\cal L}_{CG},
\label{Class-Lag0}  
\end{eqnarray}
where 
\begin{eqnarray}
{\cal L}_{WIST} &=&  \sqrt{-g} \left( \frac{1}{12} \phi^2 R + \frac{1}{2} g^{\mu\nu} \partial_\mu \phi \partial_\nu \phi \right),
\nonumber\\
{\cal L}_{CG} &=& - \sqrt{-g} \alpha_c C_{\mu\nu\rho\sigma} C^{\mu\nu\rho\sigma}.
\label{Class-Lag01}  
\end{eqnarray}
Here $\phi$ is a real scalar field with a ghost-like kinetic term, $R$ the scalar curvature, 
$\alpha_c$ a dimensionless positive coupling constant ($\alpha_c > 0$) and $C_{\mu\nu\rho\sigma}$ is conformal 
tensor defined as
\begin{eqnarray}
C_{\mu\nu\rho\sigma} &=& R_{\mu\nu\rho\sigma} - \frac{1}{2} ( g_{\mu\rho} R_{\nu\sigma}
- g_{\mu\sigma} R_{\nu\rho} - g_{\nu\rho} R_{\mu\sigma} + g_{\nu\sigma} R_{\mu\rho} )
\nonumber\\
&+& \frac{1}{6} ( g_{\mu\rho} g_{\nu\sigma} - g_{\mu\sigma} g_{\nu\rho} ) R.
\label{C-tensor}  
\end{eqnarray}

In order to perform the canonical quantization, it is more convenient to introduce an auxiliary symmetric tensor 
$K_{\mu\nu} = K_{\nu\mu}$ and a St\"{u}ckelberg-like vector field $A_\mu$,\footnote{The St\"{u}ckelberg-like 
vector field $A_\mu$ is introduced to avoid the second-class constraint.}  and rewrite ${\cal L}_{CG}$, which is the 
Lagrangian of conformal gravity, into a form \cite{Kimura1, Kimura2, Kimura3, Kubo}:
\begin{eqnarray}
{\cal L}_{CG}^{(K)} \equiv \sqrt{-g} \Bigl\{ \gamma G_{\mu\nu} K^{\mu\nu} 
+ \alpha [ ( K_{\mu\nu} - \nabla_\mu A_\nu - \nabla_\nu A_\mu )^2 
- ( K - 2 \nabla_\rho A^\rho )^2 ] \Bigr\},
\label{CG-KLag}  
\end{eqnarray}
where $G_{\mu\nu} \equiv R_{\mu\nu} - \frac{1}{2} g_{\mu\nu} R$ denotes the Einstein tensor,
and $\gamma$ and $\alpha$ are dimensionless coupling constants which obey a relation
\begin{eqnarray}
\alpha_c = \frac{\gamma^2}{8 \alpha},
\label{Couplings}  
\end{eqnarray}
where $\alpha > 0$. It is easy to see that carrying out the path integral over $K_{\mu\nu}$ in ${\cal L}_{CG}^{(K)}$ 
produces the Lagrangian of conformal gravity ${\cal L}_{CG}$. Actually, taking the variation of $K_{\mu\nu}$ 
leads to the equation
\begin{eqnarray}
K_{\mu\nu} - \nabla_\mu A_\nu - \nabla_\nu A_\mu - g_{\mu\nu} ( K - 2 \nabla_\rho A^\rho )
= - \frac{\gamma}{2 \alpha} G_{\mu\nu}.
\label{K-var}  
\end{eqnarray}
Moreover, taking the trace of this equation yields
\begin{eqnarray}
K - 2 \nabla_\rho A^\rho = - \frac{\gamma}{6 \alpha} R.
\label{K-trace}  
\end{eqnarray}
Inserting (\ref{K-trace}) to (\ref{K-var}) gives us the expression of $K_{\mu\nu}$
\begin{eqnarray}
K_{\mu\nu} = \nabla_\mu A_\nu + \nabla_\nu A_\mu - \frac{\gamma}{2 \alpha} \left( R_{\mu\nu}
- \frac{1}{6} g_{\mu\nu} R \right).
\label{K-form}  
\end{eqnarray}
Finally, substituting Eqs. (\ref{K-trace}) and (\ref{K-form}) into the Lagrangian (\ref{CG-KLag})
and using the relation (\ref{Couplings}), we can arrive at the Lagrangian of conformal gravity, ${\cal L}_{CG}$
in (\ref{Class-Lag01}) up to surface terms. This can be achieved by use of the identity
\begin{eqnarray}
C_{\mu\nu\rho\sigma}^2  = I + 2 R_{\mu\nu}^2 - \frac{2}{3} R^2,
\label{C-ident}  
\end{eqnarray}
where $I$ is defined as
\begin{eqnarray}
I = R_{\mu\nu\rho\sigma}^2 - 4 R_{\mu\nu}^2 + R^2,
\label{E-def}  
\end{eqnarray}
which is locally a total derivative in four dimensions.

From now on, as a classical Lagrangian ${\cal L}_c$ we take a linear combination of ${\cal L}_{WIST}$ 
and ${\cal L}_{CG}^{(K)}$
\begin{eqnarray}
{\cal L}_c &\equiv& {\cal L}_{WIST} + {\cal L}_{CG}^{(K)}
\nonumber\\
&=& \sqrt{-g} \Bigl\{ \frac{1}{12} \phi^2 R + \frac{1}{2} g^{\mu\nu} \partial_\mu \phi \partial_\nu \phi 
+ \gamma G_{\mu\nu} K^{\mu\nu} 
\nonumber\\
&+& \alpha [ ( K_{\mu\nu} - \nabla_\mu A_\nu - \nabla_\nu A_\mu )^2 - ( K - 2 \nabla_\rho A^\rho )^2 ] \Bigr\}.
\label{Class-Lag}  
\end{eqnarray}
The classical Lagrangian ${\cal L}_c$ is invariant under three local transformations, those are, infinitesimal 
general coordinate transformation (GCT) $\delta^{(1)}$, Weyl transformation $\delta^{(2)}$ and St\"{u}ckelberg
transformation $\delta^{(3)}$. Concretely, the GCT takes the form
\begin{eqnarray}
&{}& \delta^{(1)} g_{\mu\nu} = - ( \nabla_\mu \xi_\nu + \nabla_\nu \xi_\mu )
= - ( \xi^\alpha \partial_\alpha g_{\mu\nu} + \partial_\mu \xi^\alpha g_{\alpha\nu} 
+ \partial_\nu \xi^\alpha g_{\alpha\mu} ), 
\nonumber\\
&{}& \delta^{(1)} \phi =  - \xi^\alpha \partial_\alpha \phi,  \quad
 \delta^{(1)} K_{\mu\nu} = - \xi^\alpha \nabla_\alpha K_{\mu\nu} - \nabla_\mu \xi^\alpha K_{\alpha\nu}
- \nabla_\nu \xi^\alpha K_{\mu\alpha},
\nonumber\\
&{}& \delta^{(1)} A_\mu = - \xi^\alpha \nabla_\alpha A_\mu - \nabla_\mu \xi^\alpha A_\alpha.
\label{GCT}  
\end{eqnarray}
As for the Weyl transformation, we have
\begin{eqnarray}
&{}& \delta^{(2)} g_{\mu\nu} = 2 \Lambda g_{\mu\nu}, \quad
\delta^{(2)} \phi =  - \Lambda \phi,
\nonumber\\
&{}& \delta^{(2)} K_{\mu\nu} = \frac{\gamma}{\alpha} \nabla_\mu \nabla_\nu \Lambda 
- 2 ( A_\mu \partial_\nu \Lambda + A_\nu \partial_\mu \Lambda - g_{\mu\nu} A_\alpha \partial^\alpha \Lambda ),
\nonumber\\
&{}& \delta^{(2)} A_\mu = 0.
\label{Weyl}  
\end{eqnarray}
Note that $\delta^{(2)} K_{\mu\nu}$ has been obtained via Eq. (\ref{K-form}).
Finally, the St\"{u}ckelberg transformation is given by
\begin{eqnarray}
&{}& \delta^{(3)} g_{\mu\nu} = \delta^{(3)} \phi =  0, \quad
\delta^{(3)} K_{\mu\nu} = \nabla_\mu \varepsilon_\nu + \nabla_\nu \varepsilon_\mu, 
\nonumber\\
&{}& \delta^{(3)} A_\mu = \varepsilon_\mu.
\label{Stuckel}  
\end{eqnarray}
In the above, $\xi_\mu, \Lambda$ and $\varepsilon_\mu$ are infinitesimal transformation parameters.

To close this section, let us count the number of phyical degrees of freedom since it is known that 
this counting is more subtle in higher derivative theories than in conventional second-order derivative
theories \cite{Lee, Riegert}. In the formalism at hand, however, the introduction of the auxiliary field $K_{\mu\nu}$
makes it possible to rewrite conformal gravity with fourth-order derivatives to a second-order
derivative theory, so we can apply the usual counting method. The fields $g_{\mu\nu}, \phi, K_{\mu\nu}$ and $A_\mu$
have 10, 1, 10 and 4 degrees of freedom, respectively. We have three kinds of local symmetries, those are,
the GCT, Weyl and St\"{u}ckelberg symmetries with 4, 1 and 4 degrees of freedom, respectively. Thus,
we have totally $(10 + 1 + 10 + 4) - (4 + 1 + 4) \times 2 = 7$ physical degrees of freedom, which will turn out to
be the massless graviton of 2 physical degrees with positive-definite norm and the massive ghost of spin-2 
of 5 degrees with indefinite norm.

\section{Quantum theory}

To fix three local symmetries and obtain a BRST invariant quantum Lagrangian, we have to introduce 
three kinds of gauge fixing conditions and the corresponding Faddeev-Popov (FP) ghost terms 
in the classical Lagrangian (\ref{Class-Lag}). In our previous papers \cite{Oda-Q, Oda-W, Oda-Saake},
we have constructed two independent BRST transformations corresponding to general coordinate 
transformation (GCT) and Weyl transformation in the sense that the two nilpotent BRST charges anticommute
with each other. To do so, it has been emphasized that a gauge condition for one local symmetry must 
respect the other symmetry \cite{Oda-W}. Concretely speaking, a gauge condition for the GCT must be
invariant under the Weyl transformation while a gauge condition for the Weyl transformation must be so 
under the GCT. We would like to stress that the existence of independent BRST transformations 
makes it easy to derive many equal-time (anti)commutation relations with the help of the canonical
(anti)commutation relations and field equations as can be seen in Section 5,

However, it will turn out that we cannot find such suitable gauge fixing conditions in the present formalism 
since the gauge fixing condition for the St\"{u}ckelberg gauge transformation necessarily breaks 
the Weyl symmetry. Thus, in this article, instead of making three independent BRST charges we will construct 
only two independent BRST charges, by which physical states and observables are defined consistently. 

The suitable gauge condition for the GCT, which preserves the maximal global symmetry as will be seen later, 
is given by ``the extended de Donder gauge condition'' \cite{Oda-W}:\footnote{Let us note that this gauge condition 
breaks the general coordinate invariance, but it is invariant under the general linear transformation $GL(4)$. 
It is also straightforward to show that gauge conditions for the other local symmetries, which will be discussed
later, do not violate the $GL(4)$ symmetry. Thus, the quantum Lagrangian is also invariant under the $GL(4)$.}
\begin{eqnarray}
\partial_\mu ( \tilde g^{\mu\nu} \phi^2 ) = 0,
\label{Ext-de-Donder}  
\end{eqnarray}
where we have defined $\tilde g^{\mu\nu} \equiv \sqrt{-g} g^{\mu\nu}$. This gauge condition breaks 
the GCT (\ref{GCT}) but is invariant under both the Weyl transformation (\ref{Weyl}) and the 
St\"{u}ckelberg transformation (\ref{Stuckel}).

As the gauge fixing condition for the Weyl transformation, we shall choose, what we call, ``the traceless 
gauge condition'':\footnote{As seen in Eq. (\ref{K-trace}), this gauge condition is equivalent to the condition 
of the vanishing scalar curvature, $R = 0$ at the classical level \cite{Oda-R, Kamimura, Oda-RWS}.} 
\begin{eqnarray}
K - 2 \nabla_\mu A^\mu = 0.
\label{Traceless-gauge}  
\end{eqnarray}
Let us note that the traceless gauge condition is invariant under the GCT (\ref{GCT}) and 
the St\"{u}ckelberg transformation (\ref{Stuckel}).

Incidentally, what is called, ``the scalar gauge'':
\begin{eqnarray}
\partial_\mu ( \tilde g^{\mu\nu} \phi \partial_\nu \phi ) = 0,
\label{Scalar-gauge}  
\end{eqnarray}
which, together with the extended de Donder gauge condition (\ref{Ext-de-Donder}), assures 
the masslessness of the dilaton, turns out be inappropriate since it does not fix any dynamical degree of freedom 
associated with the gauge field $A_\mu$. 

Finally, let us consider the gauge fixing condition for the St\"{u}ckelberg transformation. It is here that
we cannot find the gauge fixing condition which breaks the St\"{u}ckelberg transformation but is invariant
under both the GCT and the Weyl transformation. Let us argue this issue in detail since this problem is interesting 
in its own right.  For instance, the gauge fixing condition for the St\"{u}ckelberg transformation which is invariant
under the GCT and the Weyl transformation would be
\begin{eqnarray}
\nabla_\mu ( \sqrt{-g} F^{\mu\nu} ) = \partial_\mu ( \sqrt{-g} F^{\mu\nu} ) = 0,
\label{Max-gauge}  
\end{eqnarray}
where $F_{\mu\nu}$ is the field strength of the St\"{u}ckelberg vector field $A_\mu$ defined as
\begin{eqnarray}
F_{\mu\nu} = \nabla_\mu A_\nu - \nabla_\nu A_\mu = \partial_\mu A_\nu - \partial_\nu A_\mu.
\label{Field-str}  
\end{eqnarray}
However, the existence of the identity $\partial_\mu \partial_\nu ( \sqrt{-g} F^{\mu\nu} ) = 0$ implies that the
gauge condition gives us only three independent equations. To supplement one more equation, we
further impose a gauge-fixing condition
\begin{eqnarray}
\nabla_\mu ( \tilde g^{\mu\nu} \phi^2 A_\nu ) = \partial_\mu ( \tilde g^{\mu\nu} \phi^2 A_\nu ) = 0.
\label{Lor-gauge}  
\end{eqnarray}
At first sight, the gauge conditions (\ref{Max-gauge}) and (\ref{Lor-gauge}) might be a suitable choice as 
the gauge condition for the St\"{u}ckelberg transformation but after integrating over the auxiliary symmetric tensor 
$K_{\mu\nu}$, we can restore the term $C_{\mu\nu\rho\sigma}^2$. This fact implies that since the St\"{u}ckelberg 
vector field $A_\mu$ plays no role in removing the second-class constraint, the gauge conditions (\ref{Max-gauge}) 
and (\ref{Lor-gauge}) do not do the job for quantizing conformal gravity properly, and we are therefore led to imposing 
the gauge condition on $K_{\mu\nu}$. Then, a natural gauge fixing condition reads
\begin{eqnarray}
\nabla_\mu K^{\mu\nu}  = 0.
\label{K-gauge}  
\end{eqnarray}
Since this gauge condition is manifestly invariant under the GCT but is not so under the Weyl transformation, 
we cannot define three independent BRST charges, but only two independent BRST charges. We will call this gauge 
condition (\ref{K-gauge}) ``the K-gauge''.

The BRST transformation corresponding to the GCT, which is called GCT BRST transformation 
$\delta^{(1)}_B$, can be obtained from (\ref{GCT}) by replacing the transformation parameter $\xi^\mu$ 
with the Faddeev-Popov (FP) ghost $c^\mu$  
\begin{eqnarray}
&{}& \delta^{(1)}_B g_{\mu\nu} = - ( \nabla_\mu c_\nu + \nabla_\nu c_\mu )
= - ( c^\alpha \partial_\alpha g_{\mu\nu} + \partial_\mu c^\alpha g_{\alpha\nu} 
+ \partial_\nu c^\alpha g_{\alpha\mu} ), 
\nonumber\\
&{}& \delta^{(1)}_B \phi =  - c^\alpha \partial_\alpha \phi, \quad 
\delta^{(1)}_B K_{\mu\nu} = - c^\alpha \nabla_\alpha K_{\mu\nu} - \nabla_\mu c^\alpha K_{\alpha\nu}
- \nabla_\nu c^\alpha K_{\mu\alpha}, 
\nonumber\\
&{}& \delta^{(1)}_B A_\mu = - c^\alpha \nabla_\alpha A_\mu - \nabla_\mu c^\alpha A_\alpha, \qquad
\delta^{(1)}_B c^\mu = - c^\alpha \partial_\alpha c^\mu,
\nonumber\\
&{}& \delta^{(1)}_B \bar c_\mu = i B_\mu, \quad
\delta^{(1)}_B B_\mu = 0, \quad
\delta^{(1)}_B b_\mu = - c^\alpha \partial_\alpha b_\mu,
\label{GCT-BRST}  
\end{eqnarray}
where $\bar c_\mu$ and $B_\mu$ are respectively an antighost and a Nakanishi-Lautrup (NL) field, and
a new NL field $b_\mu$ is defined as
\begin{eqnarray}
b_\mu = B_\mu - i c^\alpha \partial_\alpha \bar c_\mu,
\label{new-b}  
\end{eqnarray}
which will be used in place of $B_\mu$ in what follows.

On the other hand, because of the K-gauge condition (\ref{K-gauge}), in order to construct another BRST transformation which is 
independent of the GCT BRST transformation (\ref{GCT-BRST}), we make a BRST transformation in a such way that
it involves both the Weyl and the  St\"{u}ckelberg transformations simultaneously.  This new BRST transformation $\delta^{(2)}_B$, 
which we call ``WS BRST transformation'',  can be made by replacing $\Lambda$ and $\varepsilon_\mu$ with the FP ghosts $c$ 
and $\zeta_\mu$, respectively, as follows:
\begin{eqnarray}																																																						
&{}& \delta^{(2)}_B g_{\mu\nu} = 2 c g_{\mu\nu}, \quad
\delta^{(2)}_B \phi =  - c \phi, 
\nonumber\\
&{}& \delta^{(2)}_B K_{\mu\nu} = \frac{\gamma}{\alpha} \nabla_\mu \nabla_\nu c
- 2 ( A_\mu \partial_\nu c + A_\nu \partial_\mu c - g_{\mu\nu} A_\alpha \partial^\alpha c )
+ \nabla_\mu \zeta_\nu + \nabla_\nu \zeta_\mu,
\nonumber\\
&{}& \delta^{(2)}_B A_\mu = \zeta_\mu,   \quad
\delta^{(2)}_B \bar c = i B,   \quad
\delta^{(2)}_B c = \delta^{(2)}_B B  = 0,
\nonumber\\
&{}& \delta^{(2)}_B \bar \zeta_\mu = i \beta_\mu,  \quad
\delta^{(2)}_B \zeta_\mu = \delta^{(2)}_B \beta_\mu = 0
\label{WS-BRST}  
\end{eqnarray}
where $\bar c$ and $\bar \zeta_\mu$ are antighosts, and $B$ and $\beta_\mu$ are NL fields. 
In place of $\zeta_\mu$, it is more convenient to introduce a new FP ghost $\tilde \zeta_\mu$, which is defined as
\begin{eqnarray}
\tilde \zeta_\mu = \zeta_\mu + \frac{\gamma}{2 \alpha} \partial_\mu c.
\label{New-zeta}  
\end{eqnarray}
In addition to it, we introduce a new NL field $b$ which is defined as
\begin{eqnarray}
b = B + 2 i \bar c c.
\label{New-W-NL}  
\end{eqnarray}
Using the new FP ghost $\tilde \zeta_\mu$ and the new $b$ field, the WS BRST transformation for
 $K_{\mu\nu}, A_\mu, \tilde \zeta_\mu$ and $b$ can be written as
\begin{eqnarray}																																																						
&{}& \delta^{(2)}_B K_{\mu\nu} = \nabla_\mu \tilde \zeta_\nu + \nabla_\nu \tilde \zeta_\mu 
- 2 ( A_\mu \partial_\nu c + A_\nu \partial_\mu c - g_{\mu\nu} A_\alpha \partial^\alpha c ),
\nonumber\\
&{}& \delta^{(2)}_B A_\mu = \tilde \zeta_\mu - \frac{\gamma}{2 \alpha} \partial_\mu c,   \quad
\delta^{(2)}_B \tilde \zeta_\mu = 0,   \quad
\delta^{(2)}_B b = - 2 b c.
\label{zeta-WS-BRST}  
\end{eqnarray}

In order to make the two nilpotent BRST transformations be anticommutative, i.e., $\{ \delta^{(1)}_B, \delta^{(2)}_B \} = 0$, 
we must determine the remaining BRST transformations: As for the GCT BRST transformation, 
the BRST transformations on fields, which do not appear in (\ref{GCT-BRST}) 
but appear in (\ref{WS-BRST}) are determined in such a way that they coincide with their tensor structure, 
for instance,
\begin{eqnarray}
\delta^{(1)}_B c = - c^\alpha \partial_\alpha c, \qquad
\delta^{(1)}_B \tilde \zeta_\mu = - c^\alpha \nabla_\alpha \tilde \zeta_\mu - \nabla_\mu c^\alpha \tilde \zeta_\alpha.
\label{GCT-BRST2}  
\end{eqnarray}
On the other hand, in cases of the WS BRST transformations, one simply defines the vanishing BRST transformations, e.g.,
\begin{eqnarray}
\delta^{(2)}_B b_\mu = \delta^{(2)}_B c^\mu  = \delta^{(2)}_B \bar c_\mu = 0.
\label{WS-BRST2}  
\end{eqnarray}

Now that we have chosen gauge fixing conditions and established BRST transformations, we can construct
a gauge fixed and BRST invariant quantum Lagrangian by following the standard recipe:
\begin{eqnarray}
{\cal L}_q &=& {\cal L}_c + i \delta_B^{(1)} ( \tilde g^{\mu\nu} \phi^2 \partial_\mu \bar c_\nu )
+ i \delta_B^{(2)} \{ \sqrt{-g} [ \bar c ( K - 2 \nabla_\mu A^\mu ) 
+ \bar \zeta_\nu \nabla_\mu K^{\mu\nu} ] \}
\nonumber\\
&=& \sqrt{-g} \Biggl\{ \frac{1}{12} \phi^2 R + \frac{1}{2} g^{\mu\nu} \partial_\mu \phi \partial_\nu \phi 
+ \gamma G_{\mu\nu} K^{\mu\nu} + \alpha [ ( K_{\mu\nu} - \nabla_\mu A_\nu 
\nonumber\\
&-& \nabla_\nu A_\mu )^2 - ( K - 2 \nabla_\rho A^\rho )^2 ] \Biggr\}
- \tilde g^{\mu\nu} \phi^2 ( \partial_\mu b_\nu + i \partial_\mu \bar c_\lambda  \partial_\nu c^\lambda )
\nonumber\\
&-& \sqrt{-g} \, b \, ( K - 2 \nabla_\mu A^\mu ) + i \frac{\gamma}{\alpha} \tilde g^{\mu\nu} \partial_\mu \bar c \partial_\nu c
- \sqrt{-g} \nabla_\mu K^{\mu\nu} \beta_\nu 
\nonumber\\
&+& i \sqrt{-g} \nabla^\mu \bar \zeta^\nu [ \nabla_\mu \tilde \zeta_\nu
+ \nabla_\nu \tilde \zeta_\mu -  2 ( A_\mu \partial_\nu c
+ A_\nu \partial_\mu c - g_{\mu\nu} A_\alpha \partial^\alpha c ) ]
\nonumber\\
&-& i \sqrt{-g} \bar \zeta^\mu ( 2 K_{\mu\nu} \partial^\nu c - K \partial_\mu c ),
\label{q-Lag}  
\end{eqnarray}
where surface terms are dropped. 

From the Lagrangian ${\cal L}_q$, it is straightforward to derive the field equations by taking 
the variation with respect to each fundamental field in order. All the field equations are summarized
as follows:
\begin{eqnarray}
&{}& \frac{1}{12} \phi^2 G_{\mu\nu} - \frac{1}{12} ( \nabla_\mu \nabla_\nu - g_{\mu\nu} \Box ) \phi^2  
- \frac{1}{2} ( E_{\mu\nu} - \frac{1}{2} g_{\mu\nu} E )
\nonumber\\
&{}& - \frac{1}{2} g_{\mu\nu} [ \gamma G_{\rho\sigma} K^{\rho\sigma} + \alpha ( \hat K_{\rho\sigma}^2
- \hat K^2 ) ] + \gamma \Bigl[ 2 G_{\rho(\mu} K_{\nu)}\,^\rho - \nabla_\rho \nabla_{(\mu} K_{\nu)}\,^\rho
+ \frac{1}{2} \Box K_{\mu\nu}
\nonumber\\
&{}& + \frac{1}{2} g_{\mu\nu} \nabla_\rho \nabla_\sigma K^{\rho\sigma} + \frac{1}{2} K_{\mu\nu} R
- \frac{1}{2} R_{\mu\nu} K - \frac{1}{2} ( g_{\mu\nu} \Box - \nabla_\mu \nabla_\nu ) K \Bigr]
\nonumber\\
&{}& + 2 \alpha [ \hat K_{\rho(\mu} \hat K_{\nu)}\,^\rho + 2 \nabla_\rho ( A_{(\mu} \hat K_{\nu)}\,^\rho )
- \nabla_\rho ( A^\rho \hat K_{\mu\nu} ) - \hat K \hat K_{\mu\nu} - 2 \nabla_{(\mu} ( \hat K A_{\nu)} )
\nonumber\\
&{}& + g_{\mu\nu} \nabla_\rho ( \hat K A^\rho ) ] + K^\rho\,_{(\mu} \nabla_{\nu)} \beta_\rho 
+ \frac{1}{2} \nabla_\rho ( K_{\mu\nu} \beta^\rho ) - \nabla_\rho K^\rho\,_{(\mu} \beta_{\nu)}
\nonumber\\
&{}& - g_{\mu\nu} \Bigl( \nabla_\rho K^{\rho\sigma} \beta_\sigma + \frac{1}{2} K^{\rho\sigma} \nabla_\rho \beta_\sigma \Bigr)
+ L_{\mu\nu} - \frac{1}{2} g_{\mu\nu} L + N_{\mu\nu}.
\label{Field-eq1}  
\\
&{}& \frac{1}{6} \phi^2 R - E - \frac{1}{\sqrt{-g}} \partial_\mu ( \tilde g^{\mu\nu} \phi \partial_\nu \phi ) = 0.
\label{Field-eq2}
\\
&{}& \hat K_{\mu\nu} - g_{\mu\nu} \hat K = - \frac{\gamma}{2 \alpha} G_{\mu\nu} 
+ \frac{1}{2 \alpha} g_{\mu\nu} b - \frac{1}{2 \alpha} \nabla_{(\mu} \beta_{\nu)} 
+ i \frac{1}{2 \alpha} ( 2 \bar \zeta_{(\mu} \partial_{\nu)} c 
\nonumber\\
&{}& - g_{\mu\nu} \bar \zeta_\rho \partial^\rho c ).
\label{Field-eq3}
\\
&{}& \nabla^\mu ( \hat K_{\mu\nu} - g_{\mu\nu} \hat K ) = \frac{1}{2\alpha} \partial_\nu b 
+ i \frac{1}{2 \alpha} [ (\nabla_\mu \bar \zeta_\nu + \nabla_\nu \bar \zeta_\mu) \partial^\mu c 
- \nabla_\rho \bar \zeta^\rho \partial_\nu c ].
\label{Field-eq4}
\\
&{}& \partial_\mu ( \tilde g^{\mu\nu} \phi^2 ) = 0. 
\label{Field-eq5}
\\
&{}& K - 2 \nabla_\mu A^\mu = 0.
\label{Field-eq6}
\\
&{}& \nabla_\mu K^{\mu\nu} = 0.
\label{Field-eq7}
\\
&{}& \partial_\mu ( \tilde g^{\mu\nu} \phi^2 \partial_\nu c^\rho ) 
= \partial_\mu ( \tilde g^{\mu\nu} \phi^2 \partial_\nu \bar c_\rho ) = 0.
\label{Field-eq8}
\\
&{}& \partial_\mu ( \tilde g^{\mu\nu} \partial_\nu c ) = 0.
\label{Field-eq9}
\\
&{}& \frac{\gamma}{\alpha} \frac{1}{\sqrt{-g}} \partial_\mu ( \tilde g^{\mu\nu} \partial_\nu \bar c ) 
+ 2 \nabla^\nu [ ( \nabla_\mu \bar \zeta_\nu + \nabla_\nu \bar \zeta_\mu ) A^\mu - \nabla_\mu \bar \zeta^\mu A_\nu
\nonumber\\
&{}& + \bar \zeta^\mu K_{\mu\nu} 
- \frac{1}{2} \bar \zeta_\nu K ] = 0.
\label{Field-eq10}
\\
&{}& \nabla^\mu [ \nabla_\mu \tilde \zeta_\nu + \nabla_\nu \tilde \zeta_\mu -  2 ( A_\mu \partial_\nu c
+ A_\nu \partial_\mu c - g_{\mu\nu} A_\alpha \partial^\alpha c ) ]
+ 2 K_{\mu\nu} \partial^\mu c 
\nonumber\\
&{}& - K \partial_\nu c = 0.
\label{Field-eq11}  
\\
&{}& \nabla^\mu ( \nabla_\mu \bar \zeta_\nu + \nabla_\nu \bar \zeta_\mu ) = 0.
\label{Field-eq12}  
\end{eqnarray}
where we have defined the following quantities:
\begin{eqnarray}
&{}& \Box = g^{\mu\nu} \nabla_\mu \nabla_\nu,
\nonumber\\
&{}& E_{\mu\nu} = - \frac{1}{2} \partial_\mu \phi \partial_\nu \phi + \phi^2 ( \partial_\mu b_\nu 
+ i \partial_\mu \bar c_\lambda  \partial_\nu c^\lambda ) + ( \mu \leftrightarrow \nu ), 
\nonumber\\
&{}&
E = g^{\mu\nu} E_{\mu\nu},
\nonumber\\
&{}& \hat K_{\mu\nu} = K_{\mu\nu} - \nabla_\mu A_\nu - \nabla_\nu A_\mu, \quad
\hat K = g^{\mu\nu} \hat K_{\mu\nu} = K - 2 \nabla_\rho A^\rho,
\nonumber\\
&{}& L_{\mu\nu} = - b ( K_{\mu\nu} - 2 \nabla_{(\mu} A_{\nu)} ) 
- 2 \nabla_{(\mu} ( b A_{\nu)} ) + i \frac{\gamma}{\alpha} \partial_{(\mu} \bar c \partial_{\nu)} c,
\nonumber\\
&{}&
L = g^{\mu\nu} L_{\mu\nu},
\nonumber\\
&{}&
N_{\mu\nu} = - i \frac{1}{2} g_{\mu\nu} \nabla^\rho \bar \zeta^\sigma [ \nabla_\rho \tilde \zeta_\sigma
+ \nabla_\sigma \tilde \zeta_\rho - 2 ( A_\rho \partial_\sigma c + A_\sigma \partial_\rho c 
- g_{\rho\sigma} A_\gamma \partial^\gamma c ) ]
\nonumber\\
&{}&
+ i ( \nabla_{(\mu} \bar \zeta^\rho + \nabla^\rho \bar \zeta_{(\mu} ) 
[ \nabla_{|\rho|} \tilde \zeta_{\nu)} + \nabla_{\nu)} \tilde \zeta_\rho - 2 ( A_{|\rho|} \partial_{\nu)} c
+ A_{\nu)} \partial_\rho c - g_{\nu)\rho} A_\gamma \partial^\gamma c ) ] 
\nonumber\\
&{}& - i \nabla^\rho \{ \bar \zeta_{(\mu} [ \nabla_{|\rho|} \tilde \zeta_{\nu)} + \nabla_{\nu)} \tilde \zeta_\rho
- 2 ( A_{|\rho|} \partial_{\nu)} c + A_{\nu)} \partial_\rho c - g_{\nu)\rho} A_\gamma \partial^\gamma c ) ] \}
\nonumber\\
&{}& + i \nabla^\rho \{ \bar \zeta_\rho [ \nabla_\mu \tilde \zeta_\nu + \nabla_\nu \tilde \zeta_\mu
- 2 ( A_\mu \partial_\nu c + A_\nu \partial_\mu c - g_{\mu\nu} A_\gamma \partial^\gamma c ) ] \}
\nonumber\\
&{}&
- i \nabla_\rho \Bigl[ ( \nabla_{(\mu} \bar \zeta^\rho + \nabla^\rho \bar \zeta_{(\mu} ) \tilde \zeta_{\nu)}
- \frac{1}{2} ( \nabla_\mu \bar \zeta_\nu + \nabla_\nu \bar \zeta_\mu ) \tilde \zeta^\rho \Bigr]
+ i ( \nabla_{(\mu} \bar \zeta_{\nu)} A_\rho \partial^\rho c
\nonumber\\
&{}&
- \nabla_\rho \bar \zeta^\rho A_{(\mu} \partial_{\nu)} c ) - i \bar \zeta^\rho \Bigl[ 2 K_{\rho(\mu} \partial_{\nu)} c
- K_{\mu\nu} \partial_\rho c - \frac{1}{2} g_{\mu\nu} ( 2 K_{\rho\sigma} \partial^\sigma c - K \partial_\rho c ) \Bigr]
\nonumber\\
&{}&
- i \bar \zeta_{(\mu} ( 2 K_{\nu)\rho} \partial^\rho c - K \partial_{\nu)} c ).
\label{Ein&hat-K}  
\end{eqnarray}
Moreover, we have introduced symmetrization with weight one by round brackets, e.g.,
$A_{(\mu} B_{\nu)} \equiv \frac{1}{2} ( A_\mu B_\nu + A_\nu B_\mu )$.

Based on these field equations, we can write down the simpler type of equations for several fields.
First of all, using Eqs. (\ref{Field-eq5}) and (\ref{Field-eq8}), it is easy to see that
\begin{eqnarray}
g^{\mu\nu} \partial_\mu \partial_\nu c^\rho = g^{\mu\nu} \partial_\mu \partial_\nu \bar c_\rho = 0.
\label{(Anti)ghost-eq}  
\end{eqnarray}
Furthermore, taking the GCT BRST transformation of the field equation for $\bar c_\rho$ in (\ref{(Anti)ghost-eq}) 
enables us to derive the field equation for $b_\rho$ \cite{Oda-Saake}:
\begin{eqnarray}
g^{\mu\nu} \partial_\mu \partial_\nu b_\rho = 0.
\label{b-rho-eq}  
\end{eqnarray}

In other words, setting $X^M = \{ x^\mu, b_\mu, c^\mu, \bar c_\mu \}$, $X^M$ turns out to obey the very simple equation:
\begin{eqnarray}
g^{\mu\nu} \partial_\mu \partial_\nu X^M = 0.
\label{X-M-eq}  
\end{eqnarray}
This equation, together with the gauge condition $\partial_\mu ( \tilde g^{\mu\nu} \phi^2 ) = 0$, produces the two kinds 
of conserved currents:
\begin{eqnarray}
{\cal P}^{\mu M} &\equiv& \tilde g^{\mu\nu} \phi^2 \partial_\nu X^M 
= \tilde g^{\mu\nu} \phi^2 \bigl( 1 \overset{\leftrightarrow}{\partial}_\nu X^M \bigr)
\nonumber\\
{\cal M}^{\mu M N} &\equiv& \tilde g^{\mu\nu} \phi^2 \bigl( X^M 
\overset{\leftrightarrow}{\partial}_\nu Y^N \bigr),
\label{Cons-currents}  
\end{eqnarray}
where we have defined $X^M \overset{\leftrightarrow}{\partial}_\mu Y^N \equiv X^M \partial_\mu Y^N - ( \partial_\mu X^M ) Y^N$. 
Using these currents, we can show that there is a Poincar${\rm{\acute{e}}}$-like $\IOSp(8|8)$ symmetry in the present
theory as in Einstein's gravity \cite{Nakanishi, N-O-text}, which should be contrasted to the $\IOSp(10|10)$ symmetry in both Weyl invariant 
scalar-tensor gravity in Riemann geometry \cite{Oda-W} and Weyl conformal gravity in Weyl geometry \cite{Oda-Saake}. 

Here it is worth mentioning that this reduction of the global symmetry is relevant to the fact that 
the Einstein's gravity is in a sense similar to the quantum electrodynamics (QED) while the quantum conformal
gravity under consideration is similar to the quantum chromodynamics (QCD). For instance, as a representative
of the global symmetries, let us consider the BRST charges. As is well known, in the QED, the BRST charge
takes the simple form
\begin{eqnarray}
Q_B^{({\rm{QED}})} = \int d^3 x ( B \partial_0 c - \partial_0 B c ) = \int d^3 x B \overleftrightarrow{\partial_0} c,
\label{QED-BRST}  
\end{eqnarray}
where $B$ and $c$ are the NL field and antighost for the U(1) gauge symmetry, respectively.   
On the other hand, in the QCD, the BRST charge has nonlinear and interacting terms as well as quadratic ones
\begin{eqnarray}
Q_B^{({\rm{QCD}})} = \int d^3 x ( B^a D_0 c^a - \partial_0 B^a c^a + \frac{i}{2} g f_{abc} \partial_0 \bar c^a c^b c^c ),
\label{QCD-BRST}  
\end{eqnarray}
where $B^a$ and $c^a$ are the NL field and antighost for the nonabelian gauge symmetry, respectively,
$D_\mu$ the covariant derivative, $g$ the coupling constant, and $f_{abc}$ the structure constant. 
  
Analogously, in the Weyl invariant scalar-tensor gravity, the Weyl BRST charge is of the form \cite{Oda-Saake}
\begin{eqnarray}
Q_B^{({\rm{Weyl}})} = \int d^3 x \tilde g^{0\mu} \phi^2 B \overleftrightarrow{\partial_\mu} c,
\label{Weyl-BRST-charge}  
\end{eqnarray}
whereas in the quantum conformal gravity, it turns out that the WS BRST charge has a very complicated nonlinear 
structure.\footnote{Owing to its long expression, we omit to write down the charge. Even in the
Weyl limit $\zeta_\mu, \bar \zeta_\mu, \beta_\mu \rightarrow 0$, the BRST charge has interacting
terms such as $\tilde g^{0\mu} \bar c c \partial_\mu c$.}  In this sense, the quantum conformal gravity is similar to QCD
rather than the QED. Lastly, let us note that such nonlinear global symmetries cannot be described by
the generators of the Poincar${\rm{\acute{e}}}$-like $\IOSp(8|8)$ symmetry.

\section{Canonical commutation relations}

In this section, we derive the concrete expressions of canonical conjugate momenta and set up the canonical 
(anti)commutation relations (CCRs), which will be used in evaluating various equal-time (anti)commutation relations (ETCRs) 
among fundamental variables in the next section. To simplify various expressions, we obey the following abbreviations 
adopted in the textbook of Nakanishi and Ojima \cite{N-O-text}:
\begin{eqnarray}
[ A, B^\prime ] &=& [ A(x), B(x^\prime) ] |_{x^0 = x^{\prime 0}},
\qquad \delta^3 = \delta(\vec{x} - \vec{x}^\prime), 
\nonumber\\
\tilde f &=& \frac{1}{\tilde g^{00}} = \frac{1}{\sqrt{-g} g^{00}},
\label{abbreviation}  
\end{eqnarray}
where we assume that $\tilde g^{00}$ is invertible. 

To remove second order derivatives of the metric involved in $R$ and $G_{\mu\nu}$, and regard $b_\mu$
as a non-canonical variable, we perform the integration by parts and rewrite the Lagrangian 
(\ref{q-Lag}) as
\begin{eqnarray}
{\cal L}_q &=& - \frac{1}{12} \tilde g^{\mu\nu} \phi^2 ( \Gamma^\sigma_{\mu\nu} \Gamma^\alpha_{\sigma\alpha}  
-  \Gamma^\sigma_{\mu\alpha} \Gamma^\alpha_{\sigma\nu} ) 
- \frac{1}{6} \phi \partial_\mu \phi ( \tilde g^{\alpha\beta} \Gamma^\mu_{\alpha\beta}  
- \tilde g^{\mu\nu} \Gamma^\alpha_{\nu\alpha} ) 
\nonumber\\
&+& \frac{1}{2} \tilde g^{\mu\nu} \partial_\mu \phi \partial_\nu \phi 
- \gamma \sqrt{-g} ( \Gamma^\alpha_{\mu\nu} \partial_\alpha - \Gamma^\alpha_{\mu\alpha} \partial_\nu
+ \Gamma^\beta_{\mu\alpha} \Gamma^\alpha_{\beta\nu} - \Gamma^\alpha_{\mu\alpha} \Gamma^\beta_{\nu\beta} )
\bar K^{\mu\nu} 
\nonumber\\
&+& \alpha \sqrt{-g} [ ( K_{\mu\nu} - \nabla_\mu A_\nu - \nabla_\nu A_\mu )^2 - ( K - 2 \nabla_\mu A^\mu )^2 ]  
\nonumber\\
&+& \partial_\mu ( \tilde g^{\mu\nu} \phi^2 ) b_\nu - i \tilde g^{\mu\nu} \phi^2 \partial_\mu \bar c_\rho \partial_\nu c^\rho 
- \sqrt{-g} \, b \, ( K - 2 \nabla_\mu A^\mu ) + i \frac{\gamma}{\alpha} \tilde g^{\mu\nu} \partial_\mu \bar c \partial_\nu c
\nonumber\\
&-& \sqrt{-g} \nabla_\mu K^{\mu\nu} \cdot \beta_\nu 
+ i \sqrt{-g} \nabla^\mu \bar \zeta^\nu [ \nabla_\mu \tilde \zeta_\nu
+ \nabla_\nu \tilde \zeta_\mu -  2 ( A_\mu \partial_\nu c
\nonumber\\
&+& A_\nu \partial_\mu c - g_{\mu\nu} A_\alpha \partial^\alpha c ) ]
- i \sqrt{-g} \bar \zeta^\mu ( 2 K_{\mu\nu} \partial^\nu c - K \partial_\mu c )
+ \partial_\mu {\cal{V}}^\mu,
\label{Can-q-Lag}  
\end{eqnarray}
where $\bar K_{\mu\nu}$ is defined as
\begin{eqnarray}
\bar K_{\mu\nu} \equiv K_{\mu\nu} - \frac{1}{2} g_{\mu\nu} K,  \qquad
\bar K \equiv g^{\mu\nu} \bar K_{\mu\nu},
\label{bar-K}  
\end{eqnarray}
and a surface term ${\cal{V}}^\mu$ is given by
\begin{eqnarray}
{\cal{V}}^\mu &=&  \frac{1}{12} \phi^2 ( \tilde g^{\alpha\beta} \Gamma^\mu_{\alpha\beta} 
- \tilde g^{\mu\nu} \Gamma^\alpha_{\nu\alpha} ) 
+ \gamma \sqrt{-g} ( \Gamma^\mu_{\alpha\beta} \bar K^{\alpha\beta} - \Gamma^\alpha_{\alpha\nu} \bar K^{\mu\nu} )
\nonumber\\
&-& \tilde g^{\mu\nu} \phi^2 b_\nu.
\label{surface}  
\end{eqnarray}
Since the NL fields $b_\mu, b$ and $\beta_\mu$ have no derivatives in ${\cal L}_q$, we can regard them as non-canonical 
variables.

Using this Lagrangian (\ref{Can-q-Lag}), it is straightforward to derive the concrete expressions of canonical 
conjugate momenta. The result is given by
\begin{eqnarray}
\pi_g^{\mu\nu} &=& \frac{\partial {\cal L}_q}{\partial \dot g_{\mu\nu}} 
\nonumber\\
&=& - \frac{1}{24} \sqrt{-g} \phi^2 \Bigl[ - g^{0 \lambda} g^{\mu\nu} g^{\sigma\tau} - g^{0 \tau} g^{\mu\lambda} g^{\nu\sigma}
- g^{0 \sigma} g^{\mu\tau} g^{\nu\lambda} + g^{0 \lambda} g^{\mu\tau} g^{\nu\sigma} 
\nonumber\\
&+& g^{0 \tau} g^{\mu\nu} g^{\lambda\sigma} + g^{0 (\mu} g^{\nu)\lambda} g^{\sigma\tau} \Bigr] \partial_\lambda g_{\sigma\tau}
- \frac{1}{6} \sqrt{-g} \Bigl[ g^{0 (\mu} g^{\nu)\rho} - g^{\mu\nu} g^{0 \rho} \Bigr] \phi \partial_\rho \phi 
\nonumber\\
&-& \frac{1}{2} \sqrt{-g} ( 2 g^{0 (\mu} g^{\nu)\rho} -  g^{\mu\nu} g^{0 \rho} ) ( \phi^2 b_\rho + 2 b A_\rho )
\nonumber\\
&-& \gamma \sqrt{-g} \Biggl[ \nabla^{(\mu} \bar K^{\nu)0} - \frac{1}{2} \nabla^0 \bar K^{\mu\nu} 
- \frac{1}{2} g^{\mu\nu} \partial_\alpha \bar K^{0\alpha} - g^{\mu\nu} \Gamma^\beta_{\alpha\beta} \bar K^{0\alpha}
\nonumber\\
&-& 2 \Gamma^0_{\rho\sigma} g^{\rho(\mu} \bar K^{\nu)\sigma} 
+ \Gamma^\alpha_{\rho\alpha} ( g^{\rho(\mu} \bar K^{\nu)0} + g^{0(\mu} \bar K^{\nu)\rho} )  \Biggr]
\nonumber\\
&+& 2 \alpha \sqrt{-g} \Bigl[ 2 \hat K^{0(\mu} A^{\nu)}  - \hat K^{\mu\nu} A^0 
- \hat K ( 2 g^{0(\mu} A^{\nu)} - g^{\mu\nu} A^0 ) \Bigr]
\nonumber\\
&+& \frac{1}{2} \sqrt{-g} ( 2 g^{0 (\mu} K^{\nu)\rho} + g^{0\rho} K^{\mu\nu} - g^{\mu\nu} K^{0 \rho} ) \beta_\rho,
\nonumber\\
&-& i \sqrt{-g} ( 2 g^{0\alpha} g^{\beta(\mu} \bar \zeta^{\nu)} - \bar \zeta^0 g^{\alpha\mu} g^{\beta\nu} )
( \nabla_{(\alpha} \tilde \zeta_{\beta)} - 2 A_{(\alpha} \partial_{\beta)} c + g_{\alpha\beta} A_\gamma \partial^\gamma c )
\nonumber\\
&-& i \sqrt{-g} [ ( \nabla^0 \bar \zeta^{(\mu} + \nabla^{(\mu} \bar \zeta^{|0|} ) \tilde \zeta^{\nu)} 
- \nabla^{(\mu} \bar \zeta^{\nu)} \tilde \zeta^0 ],
\nonumber\\
\pi_\phi &=& \frac{\partial {\cal L}_q}{\partial \dot \phi} = \tilde g^{0 \mu} \partial_\mu \phi + 2 \tilde g^{0 \mu} \phi b_\mu
+ \frac{1}{6} \phi ( - \tilde g^{\alpha\beta} \Gamma^0_{\alpha\beta} + \tilde g^{0 \alpha} \Gamma^\beta_{\alpha\beta} ),
\nonumber\\
\pi_K^{\mu\nu} &=& \frac{\partial {\cal L}_q}{\partial \dot{K}_{\mu\nu}} 
= - \gamma \sqrt{-g} \Bigl[ ( g^{\mu\rho} g^{\nu\sigma} - \frac{1}{2} g^{\mu\nu} g^{\rho\sigma} ) 
\Gamma^0_{\rho\sigma}
- \frac{1}{2} ( g^{0\mu} g^{\nu\rho} + g^{0\nu} g^{\mu\rho} - g^{\mu\nu} g^{0\rho} ) \Gamma^\sigma_{\rho\sigma} \Bigr]
\nonumber\\ 
&-& \frac{1}{2} \sqrt{-g} ( g^{0\mu} \beta^\nu + g^{0\nu} \beta^\mu ),
\nonumber\\ 
\pi_A^\mu &=& \frac{\partial {\cal L}_q}{\partial \dot{A}_\mu} = - 4 \alpha \sqrt{-g} ( \hat K^{0\mu} - g^{0\mu} \hat K )
+ 2 \tilde g^{0\mu} b,
\nonumber\\
\pi_{c \mu} &=& \frac{\partial {\cal L}_q}{\partial \dot c^\mu} = - i \tilde g^{0 \nu} \phi^2 \partial_\nu \bar c_\mu,  \quad
\pi_{\bar c}^\mu = \frac{\partial {\cal L}_q}{\partial \dot {\bar c}_\mu} = i \tilde g^{0 \nu} \phi^2 \partial_\nu c^\mu,
\nonumber\\
\pi_c &=& \frac{\partial {\cal L}_q}{\partial \dot c} = i \frac{\gamma}{\alpha} \tilde g^{0 \mu} \partial_\mu \bar c
- 2 i \sqrt{-g} [ ( \nabla^0 \bar \zeta^\mu + \nabla^\mu \bar \zeta^0 ) A_\mu - \nabla_\rho \bar \zeta^\rho A^0 ]
- i \sqrt{-g} ( 2 \bar \zeta^\mu K_\mu \,^0 - \bar \zeta^0 K ),  
\nonumber\\
\pi_{\bar c} &=& \frac{\partial {\cal L}_q}{\partial \dot {\bar c}} = - i \frac{\gamma}{\alpha} \tilde g^{0 \mu} \partial_\mu c, 
\nonumber\\
\pi_{\tilde \zeta}^\mu &=& \frac{\partial {\cal L}_q}{\partial \dot {\tilde \zeta}_\mu} = i \sqrt{-g} ( \nabla^\mu \bar \zeta^0
+ \nabla^0 \bar \zeta^\mu ), 
\nonumber\\
\pi_{\bar \zeta}^\mu &=& \frac{\partial {\cal L}_q}{\partial \dot {\bar \zeta}_\mu} = - i \sqrt{-g} [ \nabla^\mu \tilde \zeta^0
+ \nabla^0 \tilde \zeta^\mu - 2 ( A^\mu \partial^0 c + A^0 \partial^\mu c - g^{0\mu} A_\rho \partial^\rho c ) ],
\label{CCM}  
\end{eqnarray}
where we have defined the time derivative such as $\dot g_{\mu\nu} \equiv \frac{\partial g_{\mu\nu}}{\partial t}
\equiv \frac{\partial g_{\mu\nu}}{\partial x^0} \equiv \partial_0 g_{\mu\nu}$, and differentiation of ghosts is taken 
from the right.  

Next let us set up the canonical (anti)commutation relations: 
\begin{eqnarray}
&{}& [ g_{\mu\nu}, \pi_g^{\rho\lambda\prime} ] = [ K_{\mu\nu}, \pi_K^{\rho\lambda\prime} ] 
= i \frac{1}{2} ( \delta_\mu^\rho\delta_\nu^\lambda 
+ \delta_\mu^\lambda\delta_\nu^\rho) \delta^3,   
\nonumber\\
&{}& [ \phi, \pi_\phi^\prime ] = i \delta^3,  \quad
[ A_\mu, \pi_A^{\nu\prime} ] = i \delta_\mu^\nu \delta^3, 
\nonumber\\
&{}& \{ c^\mu, \pi_{c \nu}^\prime \} = \{ \bar c_\nu, \pi_{\bar c}^{\mu\prime} \}
= i \delta_\nu^\mu \delta^3,    \quad
\{ c, \pi_c^\prime  \} = \{ \bar c, \pi_{\bar c}^\prime \} = i \delta^3,
\nonumber\\
&{}& \{ \bar{\zeta}_\mu, \pi_{\bar \zeta}^{\nu\prime} \} 
= \{ \tilde \zeta_\mu, \pi_{\tilde \zeta}^{\nu\prime} \} = i \delta_\mu^\nu \delta^3,  
\label{CCRs}  
\end{eqnarray}
where the other (anti)commutation relations vanish. 
In setting up these CCRs, it is valuable to distinguish non-canonical variables from canonical ones. 
Recall again that in our formalism, the NL fields $b_\mu, b$ and $\beta_\mu$ are not canonical variables.

\section{Equal-time commutation relations}

Since we have presented the canonical (anti)commutation relations (CCRs) in the previous section, 
we would like to evaluate various nontrivial equal-time (anti)commutation relations (ETCRs) 
which are necessary for the algebra of symmetries and computations in later sections.   
In what follows, we will derive various important equal-time (anti)commutation relations (ETCRs) 
on the basis of the canonical (anti)commutation relations, field equations and BRST transformations. 
In deriving ETCRs, we often use a useful identity for generic variables $\Phi$ and $\Psi$
\begin{eqnarray}
[ \Phi, \dot \Psi^\prime] = \partial_0 [ \Phi, \Psi^\prime] - [ \dot \Phi, \Psi^\prime],
\label{identity}  
\end{eqnarray}
which holds for the anticommutation relation as well. 

To begin with, we wish to derive the ETCR between $g_{\mu\nu}$ and $b_\mu$, which is one of 
the important ETCRs and plays a role in proving the algebra of symmetries. For this purpose,
let us first consider the antiCCR, $\{ c^\mu, \pi_{c\nu}^\prime \} = i \delta_\nu^\mu \delta^3$,
which gives us 
\begin{eqnarray}
\{ c^\mu, \dot{\bar c}_\nu^\prime \} = - \tilde f \phi^{-2} \delta_\nu^\mu \delta^3.
\label{c-cbar-GCT}  
\end{eqnarray}
Next, we find that the CCR, $[ g_{\mu\nu}, \pi_{c\rho}^\prime ] = 0$ leads to
\begin{eqnarray}
[ \dot g_{\mu\nu}, \bar c_\rho^\prime ] = 0,
\label{g-cbar-GCT}  
\end{eqnarray}
where we have used the CCR, $[ g_{\mu\nu}, \bar c_\rho^\prime ] = 0$ and the formula (\ref{identity}).
It then turns out that the GCT BRST transformation (\ref{GCT-BRST}) of the CCR, $[ g_{\mu\nu}, 
\bar c_\rho^\prime ] = 0$ yields
\begin{eqnarray}
[ g_{\mu\nu}, b_\rho^\prime ] = - i \tilde f \phi^{-2} ( \delta_\mu^0 g_{\rho\nu} + \delta_\nu^0 g_{\rho\mu} ) \delta^3,
\label{g-b}  
\end{eqnarray}
where we have used Eqs. (\ref{new-b}), (\ref{c-cbar-GCT}) and (\ref{g-cbar-GCT}).

From this ETCR, we can easily derive ETCRs:\footnote{The latter ETCR gives us 
$[ \phi^2 b_\mu, \tilde g^{0\nu\prime} ] = - i \delta_\mu^\nu \delta^3$, which implies that $\phi^2 b_\mu$
corresponds to the canonical conjugate momentum of $\tilde g^{0\mu}$.}
\begin{eqnarray}
&{}& [ g^{\mu\nu}, b_\rho^\prime ] = i \tilde f \phi^{-2} ( g^{\mu0} \delta_\rho^\nu+ g^{\nu0} \delta_\rho^\mu) \delta^3,
\nonumber\\
&{}& [ \tilde g^{\mu\nu}, b_\rho^\prime ] = i \tilde f \phi^{-2} ( \tilde g^{\mu0} \delta_\rho^\nu+ \tilde g^{\nu0} \delta_\rho^\mu 
- \tilde g^{\mu\nu} \delta_\rho^0 ) \delta^3.
\label{3-g-b}  
\end{eqnarray}
Here we have used the following fact; since a commutator works as a derivation, we can have formulae:
\begin{eqnarray}
&{}& [ g^{\mu\nu}, \Phi^\prime ] = - g^{\mu\alpha} g^{\nu\beta} [ g_{\alpha\beta}, \Phi^\prime ],
\nonumber\\
&{}& [ \tilde g^{\mu\nu}, \Phi^\prime ] = - \left( \tilde g^{\mu\alpha} g^{\nu\beta} - \frac{1}{2} \tilde g^{\mu\nu} 
g^{\alpha\beta} \right) [ g_{\alpha\beta}, \Phi^\prime ],
\label{Simple formulae}  
\end{eqnarray}
where $\Phi$ is a generic field. 

Now we would like to derive another important ETCR, $[ \dot g_{\rho\sigma}, g_{\mu\nu}^\prime ]$.
To this aim, let us focus on the canonical conjugate momentum $\pi_K^{\mu\nu}$, from which
we can describe $\dot g_{ij}$ as
\begin{eqnarray}
\dot g_{ij} &=& \frac{2}{\gamma} \tilde f [ ( g_{i\mu} g_{j\nu} - \frac{1}{2} g_{ij} g_{\mu\nu} ) 
\pi_K^{\mu\nu} - \frac{1}{2} \tilde g_{ij} \beta^0 ]
+ \tilde f [ \tilde g^{0\alpha} ( \partial_i g_{j\alpha} + \partial_j g_{i\alpha} ) 
\nonumber\\
&-& \tilde g^{0k} \partial_k g_{ij} ].
\label{dot-g-piK}  
\end{eqnarray}
This expression immediately gives us the ETCR
\begin{eqnarray}
[ \dot g_{ij}, g_{\mu\nu}^\prime ] = 0.
\label{g-dot-g}  
\end{eqnarray}
Here we have used the ETCR
\begin{eqnarray}
[ \beta_\rho, g_{\mu\nu}^\prime ] = 0,
\label{beta-g}  
\end{eqnarray}
which can be easily shown by taking the WS BRST transformation (\ref{WS-BRST})
of the CCR, $[ \bar \zeta_\rho, g_{\mu\nu}^\prime ] = 0$. 

In order to calculate the remaining ETCR, $[ \dot g_{0\rho}, g_{\mu\nu}^\prime ]$, we utilize 
the extended de Donder gauge condition (\ref{Ext-de-Donder}), from which we can obtain the equation:
\begin{eqnarray}
g^{\rho\sigma} \Gamma^\mu_{\rho\sigma} = 2 \phi^{-1} g^{\mu\nu} \partial_\nu \phi.
\label{Ed-Donder}  
\end{eqnarray}
This equation makes it possible to express $\dot g_{0\rho}$ in terms of $\dot g_{kl}$ and $\dot \phi$ as follows:
\begin{eqnarray}
\dot g_{00} &=& \frac{1}{g^{00}} g^{ij} \dot g_{ij} + \frac{4}{g^{00}} \phi^{-1} \dot \phi + \dots,
\nonumber\\
\dot g_{0i} &=& - \frac{1}{g^{00}} g^{0j} \dot g_{ij} + \dots,
\label{dot-g-0rho}  
\end{eqnarray}
where the ellipsis denotes terms without time-derivatives. Then, using Eq. (\ref{g-dot-g}), we find that $[ \dot g_{0i}, g_{\mu\nu}^\prime ] = 0$ and
\begin{eqnarray}
[ \dot g_{00}, g_{\mu\nu}^\prime ] = \frac{4}{g^{00}} \phi^{-1} [ \dot \phi, g_{\mu\nu}^\prime ].
\label{dot-g-00}  
\end{eqnarray}
To evaluate the right-hand side (RHS), we need to use the canonical conjugate momentum $\pi_\phi$,
from which we can express $\dot \phi$ in terms of $\pi_\phi, b_\rho$ and $\dot g_{ij}$ as 
\begin{eqnarray}
\dot \phi &=& \tilde f \Bigl\{ \pi_\phi - \tilde g^{0i} \partial_i \phi - 2 \tilde g^{0\rho} \phi b_\rho
- \frac{1}{6} \phi [ ( - \tilde g^{0i} g^{0j} + \tilde g^{00} g^{ij} ) \dot g_{ij} 
\nonumber\\
&+& ( - \tilde g^{0\rho} g^{i\sigma} + \tilde g^{0i} g^{\rho\sigma} ) \partial_i g_{\rho\sigma} ] \Bigr\}.    
\label{pi-phi-express}  
\end{eqnarray}
Then, with the help of Eqs. (\ref{g-b}) and (\ref{g-dot-g}),  Eq. (\ref{pi-phi-express}) enables us to evaluate the ETCR, $[ \dot \phi, g_{\mu\nu}^\prime ]$ 
to be
\begin{eqnarray}
[ \dot \phi, g_{\mu\nu}^\prime ] = - 4 i \tilde f^2 \sqrt{-g} \phi^{-1} \delta_\mu^0 \delta_\nu^0 \delta^3.
\label{dot-phi-g-ETCR}  
\end{eqnarray}
From Eqs. (\ref{dot-g-00}) and (\ref{dot-phi-g-ETCR}), we have 
\begin{eqnarray}
[ \dot g_{00}, g_{\mu\nu}^\prime ] = - 16 i \frac{1}{(g^{00})^2} \tilde f \phi^{-2} \delta_\mu^0 \delta_\nu^0 \delta^3.
\label{dot-g-00-2}  
\end{eqnarray}
Hence, we can arrive at the result 
\begin{eqnarray}
[ \dot g_{\rho\sigma}, g_{\mu\nu}^\prime ] = - 16 i \frac{1}{(g^{00})^2} \tilde f \phi^{-2} \delta_\rho^0 \delta_\sigma^0
\delta_\mu^0 \delta_\nu^0 \delta^3.
\label{dot-g-g-final}  
\end{eqnarray}

Incidentally, we can also offer a different proof of Eq. (\ref{dot-g-g-final}) on the basis of symmetry of this ETCR.
The ETCR, $[ \dot g_{\rho\sigma}, g_{\mu\nu}^\prime ]$ has in general a symmetry under the simultaneous exchange 
of $(\mu\nu) \leftrightarrow (\rho\sigma)$ and primed $\leftrightarrow$ unprimed in addition to the usual symmetry 
$\mu \leftrightarrow \nu$ and $\rho \leftrightarrow \sigma$.  We can therefore write down its general expression
\begin{eqnarray}
[ \dot g_{\rho\sigma}, g_{\mu\nu}^\prime ] &=& \biggl\{ c_1 g_{\rho\sigma} g_{\mu\nu} + c_2 ( g_{\rho\mu} g_{\sigma\nu}
+ g_{\rho\nu} g_{\sigma\mu} )
\nonumber\\
&+& \sqrt{-g} \tilde f \Bigl[ c_3 ( \delta_\rho^0 \delta_\sigma^0 g_{\mu\nu} + \delta_\mu^0 \delta_\nu^0 g_{\rho\sigma} )
+ c_4 ( \delta_\rho^0 \delta_\mu^0 g_{\sigma\nu} + \delta_\rho^0 \delta_\nu^0 g_{\sigma\mu} 
\nonumber\\
&+& \delta_\sigma^0 \delta_\mu^0 g_{\rho\nu} + \delta_\sigma^0 \delta_\nu^0 g_{\rho\mu} ) \Bigr]
+ ( \sqrt{-g} \tilde f )^2 c_5 \delta_\rho^0 \delta_\sigma^0 \delta_\mu^0 \delta_\nu^0 \biggr\} \delta^3,  
\label{dot-G-G}  
\end{eqnarray}
where $c_i ( i = 1, \cdots, 5)$ are some coefficients. To fix the coefficients $c_i$, let us make use of the 
extended de Donder gauge condition (\ref{Ext-de-Donder}), which can be rewritten as
\begin{eqnarray}
( g^{0\lambda} g^{\rho\sigma} - 2 g^{\lambda\rho} g^{0\sigma} ) \dot g_{\rho\sigma}
+ 4 \phi^{-1} g^{\lambda\rho} \partial_\rho \phi
= ( 2 g^{\lambda\rho} g^{\sigma i} - g^{\rho\sigma} g^{\lambda i} ) \partial_i g_{\rho\sigma}.
\label{Ext-de-Donder-g1}  
\end{eqnarray}
Using (\ref{dot-phi-g-ETCR}), Eq. (\ref{Ext-de-Donder-g1}) yields
\begin{eqnarray}
( g^{0\lambda} g^{\rho\sigma} - 2 g^{\lambda\rho} g^{0\sigma} ) [ \dot g_{\rho\sigma}, g_{\mu\nu}^\prime ] 
= 16 i \frac{1}{g^{00}} \tilde f \phi^{-2} g^{0\lambda} \delta_\mu^0 \delta_\nu^0 \delta^3.
\label{Ext-de-Donder-g2}  
\end{eqnarray}
It then turns out that this equation provides us with relations among the coefficients:
\begin{eqnarray}
c_3 = 2 ( c_1 + c_2 ), \qquad c_4 = - c_2, \qquad 2 c_3 - c_5 = 16 i \tilde f \phi^{-2}.
\label{c-relation}  
\end{eqnarray}
To fix the coefficients completely, we further take account of the CCR, $[ \pi_K^{\alpha\beta}, g_{\mu\nu}^\prime ] = 0$,
which can be cast to the form
\begin{eqnarray}
[ g^{0\alpha} ( g^{\beta\sigma} g^{0\rho} - g^{0\beta} g^{\rho\sigma} ) 
+ g^{\alpha\rho} ( g^{0\beta} g^{0\sigma} - g^{00} g^{\beta\sigma} ) ]
[ \dot g_{\rho\sigma}, g_{\mu\nu}^\prime ] = 0,
\label{pi-K-g}  
\end{eqnarray}
where Eq. (\ref{beta-g}) was used.
Substituting (\ref{dot-G-G}) into (\ref{pi-K-g}) leads to
\begin{eqnarray}
c_1 = c_2 = c_3 = 0.
\label{c_i=0}  
\end{eqnarray}
Together with Eq. (\ref{c-relation}), Eq. (\ref{c_i=0}) gives us all the vanishing coefficients $c_i = 0$ except for
$c_5 = - 16 i \tilde f \phi^{-2}$, so we have succeeded in proving the ETCR (\ref{dot-g-g-final}) again.

Next, we wish to calculate the ETCRs involving the $b_\mu$ field.    
First of all, we will show that 
\begin{eqnarray}
[ \phi, b_\rho^\prime ] = 0.
\label{phi&b}  
\end{eqnarray}
This ETCR can be obtained by using the CCRs
\begin{eqnarray}
[ \phi, \bar c_\rho^\prime ] = [ \phi, \pi_{c\rho}^\prime ] = 0.
\label{phi&b2}  
\end{eqnarray}
and the GCT BRST transformation. In fact, the two CCRs in (\ref{phi&b2}) provide the ETCRs
\begin{eqnarray}
[ \phi, \dot{\bar c}_\rho^\prime ] = [ \dot \phi, \bar c_\rho^\prime ] = 0.
\label{phi&b3}  
\end{eqnarray}
It is easy to see that with the help of Eq. (\ref{phi&b3}), the GCT BRST transformation of the former CCR 
in (\ref{phi&b2}) leads to Eq. (\ref{phi&b}).
By means of the same method, we can also show that 
\begin{eqnarray}
[ \Phi, b_\rho^\prime ] = 0,
\label{ghosts-b-rho}  
\end{eqnarray}
where $\Phi \equiv \{ c^\mu, \bar c_\mu, c, \bar c \}$. 

Similarly, we can calculate $[ K_{\mu\nu}, b_\rho^\prime ]$ as follows: Starting with the CCRs 
\begin{eqnarray}
[ K_{\mu\nu}, \bar c_\rho^\prime ] = [ K_{\mu\nu}, \pi_{c\rho}^\prime ] = 0,
\label{K&bar-c}  
\end{eqnarray}
we can obtain that
\begin{eqnarray}
[ K_{\mu\nu}, \dot{\bar c}_\rho^\prime ] = [ \dot K_{\mu\nu}, \bar c_\rho^\prime ] = 0.
\label{K&bar-c2}  
\end{eqnarray}
Taking the GCT BRST transformation of the former equation in (\ref{K&bar-c}), we have
\begin{eqnarray}
- \{ c^\alpha \nabla_\alpha K_{\mu\nu} + \nabla_\mu c^\alpha K_{\alpha\nu} 
+ \nabla_\nu c^\alpha K_{\mu\alpha}, \bar c_\rho^\prime \}
+ [ K_{\mu\nu}, i B_\rho^\prime ] = 0.
\label{K&bar-c3}  
\end{eqnarray}
Using Eqs. (\ref{c-cbar-GCT}), (\ref{K&bar-c2}) and $[ \dot g_{\mu\nu}, \bar c_\rho^\prime ] = 0$, which is easily proved,  
we reach the ETCR between $K_{\mu\nu}$ and $b_\rho$:
\begin{eqnarray}
[ K_{\mu\nu}, b_\rho^\prime ] = - i \tilde f \phi^{-2} ( \delta_\mu^0 K_{\rho\nu} + \delta_\nu^0 K_{\rho\mu} ) \delta^3.
\label{K-b-rho}  
\end{eqnarray}
Along the same line of argument, we can prove that 
\begin{eqnarray}
&{}& [ A_\mu, b_\rho^\prime ] = - i \tilde f \phi^{-2} \delta_\mu^0 A_\rho \delta^3,  \qquad
[ b, b_\rho^\prime ] = 0,  \qquad
[ \beta_\mu, b_\rho^\prime ] = - i \tilde f \phi^{-2} \delta_\mu^0 \beta_\rho \delta^3,
\nonumber\\
&{}& [ \zeta_\mu, b_\rho^\prime ] = - i \tilde f \phi^{-2} \delta_\mu^0 \zeta_\rho \delta^3,  \qquad
[ \bar \zeta_\mu, b_\rho^\prime ] = - i \tilde f \phi^{-2} \delta_\mu^0 \bar \zeta_\rho \delta^3.
\label{Var-b-rho}  
\end{eqnarray}

Now that we have established the type of the ETCRs, $[ \Phi, b_\rho^\prime ]$ with $\Phi$ being several
fields, we wish to evaluate the ETCRs with the form of $[ \dot \Phi, b_\rho^\prime ]$. 
First of all, the ETCR, $[ \dot g_{\mu\nu}, b_\rho^\prime ]$ has been already calculated by using the method 
developed in our previous article \cite{Oda-Q}. Only the result is written out as
\begin{eqnarray}
[ \dot g_{\mu\nu}, b_\rho^\prime ] &=& - i \Bigl\{ \tilde f \phi^{-2} ( \partial_\rho g_{\mu\nu} 
+ \delta_\mu^0 \dot g_{\rho\nu} + \delta_\nu^0 \dot g_{\rho\mu} ) \delta^3 
\nonumber\\
&+& [ ( \delta_\mu^k - 2 \delta_\mu^0 \tilde f \tilde g^{0 k} ) g_{\rho\nu}
+ (\mu\leftrightarrow \nu) ] \partial_k ( \tilde f \phi^{-2} \delta^3 ) \Bigr\},
\label{dot g-b}  
\end{eqnarray}
or equivalently,
\begin{eqnarray}
[ g_{\mu\nu}, \dot b_\rho^\prime ] &=& i \Bigl\{ [ \tilde f \phi^{-2} \partial_\rho g_{\mu\nu} 
- \partial_0 (\tilde f \phi^{-2} ) ( \delta_\mu^0 g_{\rho\nu} + \delta_\nu^0 g_{\rho\mu} ) ] \delta^3 
\nonumber\\
&+& [ ( \delta_\mu^k - 2 \delta_\mu^0 \tilde f \tilde g^{0k} ) g_{\rho\nu} + (\mu \leftrightarrow \nu) ]
\partial_k (\tilde f \phi^{-2} \delta^3) \Bigr\}.
\label{g-dot b}  
\end{eqnarray}

Next, let us evaluate $[ \dot \phi, b_\rho^\prime ]$. The CCR, $[ \phi, \pi_{c \rho}^\prime ] = 0$ leads to the ETCR
\begin{eqnarray}
[ \phi, \dot{\bar c}_\rho^\prime ] = 0.  
\label{phi-dot-bar-c-rho}  
\end{eqnarray}
Taking the GCT BRST transformation of this equation gives us
\begin{eqnarray}
\{ - c^\alpha \partial_\alpha \phi, \dot{\bar c}_\rho^\prime \} + [ \phi, i \dot B_\rho^\prime ] = 0.  
\label{phi-dot-bar-c-rho2}  
\end{eqnarray}
Here note that the second term on the LHS can be simplified to 
\begin{eqnarray}
[ \phi, i \dot B_\rho^\prime ] &=& [ \phi, i \dot b_\rho^\prime ] 
- [ \phi, \frac{d}{dt} ( c^{\alpha\prime} \partial_\alpha \bar c_\rho^\prime ) ]
\nonumber\\
&=&  [ \phi, i \dot b_\rho^\prime ],
\label{phi-dot-bar-c-rho3}  
\end{eqnarray}
where we have used Eq. (\ref{phi-dot-bar-c-rho}) and the ETCRs
\begin{eqnarray}
[ \phi, \dot c^{\rho\prime} ] = [ \phi, \ddot{\bar c}_\rho^\prime ] = 0.  
\label{phi-dot-bar-c-rho4}  
\end{eqnarray}
Note that the latter equation can be derived the field equation $g^{\mu\nu} \partial_\mu \partial_\nu 
\bar c_\rho = 0$ and Eq. (\ref{phi-dot-bar-c-rho}). Then, it is easy to derive the ETCR
\begin{eqnarray}
[ \dot \phi, b_\rho^\prime ] = - i \tilde f \phi^{-2} \partial_\rho \phi \delta^3.  
\label{dot-phi-b-rho}  
\end{eqnarray}
   
Furthermore, following the derivation in Appendix A, we find that 
\begin{eqnarray}
[ \dot A_\mu, b_\rho^\prime ] = i ( 2 \tilde f \tilde g^{0i} \delta_\mu^0 - \delta_\mu^i ) 
A_\rho \partial_i ( \tilde f \phi^{-2} \delta^3 ) 
- i \tilde f \phi^{-2} ( \partial_\rho A_\mu + \delta_\mu^0 \partial_0 A_\rho ) \delta^3.  
\label{dot-Amu-b}  
\end{eqnarray}
In a similar manner, we can also show that
\begin{eqnarray}
&{}& [ \dot K_{\mu\nu}, b_\rho^\prime ] = [ i ( 2 \tilde f \tilde g^{0i} \delta_\mu^0 - \delta_\mu^i ) 
K_{\rho\nu} + (\mu \leftrightarrow \nu) ] \partial_i ( \tilde f \phi^{-2} \delta^3 ) 
\nonumber\\
&{}& - i \tilde f \phi^{-2} ( \partial_\rho K_{\mu\nu} + \delta_\mu^0 \partial_0 K_{\rho\nu} 
+ \delta_\nu^0 \partial_0 K_{\rho\mu} ) \delta^3,
\nonumber\\
&{}& [ \dot b, b_\rho^\prime ] =  - i \tilde f \phi^{-2} \partial_\rho b \delta^3, \quad
[ \dot c^\mu, b_\rho^\prime ] =  - i \tilde f \phi^{-2} \partial_\rho c^\mu \delta^3, \quad
[ \dot{\bar c}_\mu, b_\rho^\prime ] =  - i \tilde f \phi^{-2} \partial_\rho \bar c_\mu \delta^3,
\nonumber\\
&{}& 
[ \dot c, b_\rho^\prime ] =  - i \tilde f \phi^{-2} \partial_\rho c \delta^3, \quad
[ \dot{\bar c}, b_\rho^\prime ] =  - i \tilde f \phi^{-2} \partial_\rho \bar c \delta^3,
\nonumber\\
&{}& [ \dot \beta_\mu, b_\rho^\prime ] = i ( 2 \tilde f \tilde g^{0i} \delta_\mu^0 - \delta_\mu^i ) 
\beta_\rho \partial_i ( \tilde f \phi^{-2} \delta^3 ) 
- i \tilde f \phi^{-2} ( \partial_\rho \beta_\mu + \delta_\mu^0 \partial_0 \beta_\rho ) \delta^3,
\nonumber\\
&{}& [ \dot \zeta_\mu, b_\rho^\prime ] = i ( 2 \tilde f \tilde g^{0i} \delta_\mu^0 - \delta_\mu^i ) 
\zeta_\rho \partial_i ( \tilde f \phi^{-2} \delta^3 ) 
- i \tilde f \phi^{-2} ( \partial_\rho \zeta_\mu + \delta_\mu^0 \partial_0 \zeta_\rho ) \delta^3,
\nonumber\\
&{}& [ \dot {\bar \zeta}_\mu, b_\rho^\prime ] = i ( 2 \tilde f \tilde g^{0i} \delta_\mu^0 - \delta_\mu^i ) 
\bar \zeta_\rho \partial_i ( \tilde f \phi^{-2} \delta^3 ) 
- i \tilde f \phi^{-2} ( \partial_\rho \bar \zeta_\mu + \delta_\mu^0 \partial_0 \bar \zeta_\rho ) \delta^3.  
\label{dot-vec1-b}  
\end{eqnarray}
Following our previous calculation \cite{Oda-Q}, we can prove
\begin{eqnarray}
&{}& [ b_\mu, b_\nu^\prime ] = 0,  
\nonumber\\
&{}& [ b_\mu, \dot b_\nu^\prime ] = i \tilde f \phi^{-2} ( \partial_\mu b_\nu + \partial_\nu b_\mu ) \delta^3. 
\label{b-b}  
\end{eqnarray}

Finally, let us evaluate $[ \dot g_{\rho\sigma}, K_{\mu\nu}^\prime ]$, which is needed for later calculations. 
Let us start with the CCR, $[ \bar \zeta_\rho, K_{\mu\nu}^\prime ] = 0$. The WS BRST transformation $\delta_B^{(2)}$
of this CCR reads
\begin{eqnarray}
[ \beta_\rho, K_{\mu\nu}^\prime ] = - i \{ \bar \zeta_\rho, \nabla_\mu \tilde \zeta_\nu^\prime
+ \nabla_\nu \tilde \zeta_\mu^\prime \},
\label{beta-rho-K}  
\end{eqnarray}
where we have used the ETCR
\begin{eqnarray}
\{ \bar \zeta_\rho, \dot c^\prime \} = 0,
\label{bar-zeta-dot-c}  
\end{eqnarray}
which can be easily obtained from the CCR, $\{ \bar \zeta_\rho, \pi_{\bar c}^\prime \} = 0$.

To calculate the right-hand side (RHS) of Eq. (\ref{beta-rho-K}), let us first calculate the ETCR, 
$[ g_{\mu\nu}, \dot{\bar \zeta}_\rho^\prime ]$. The CCR, $[ g_{\mu\nu}, \pi_{\tilde \zeta}^{\rho\prime} ] = 0$
leads to the equation
\begin{eqnarray}
( g^{0\rho\prime} g^{0\beta\prime} + g^{00\prime} g^{\rho\beta\prime} ) [ g_{\mu\nu}, \dot{\bar \zeta}_\beta^\prime ]
= ( g^{\rho\alpha\prime} g^{0\beta\prime} + g^{0\alpha\prime} g^{\rho\beta\prime} ) 
[ g_{\mu\nu}, \Gamma_{\alpha\beta}^{\gamma\prime} ] \bar \zeta_\gamma^\prime.
\label{g-bar-zeta-Gamma}  
\end{eqnarray}
Since from Eq. (\ref{dot-g-g-final}) we can derive the ETCR
\begin{eqnarray}
[ g_{\mu\nu}, \Gamma_{\alpha\beta}^{\gamma\prime} ] = - 8 i \frac{1}{(g^{00})^2} \tilde f \phi^{-2}
g^{0\gamma} \delta_\mu^0 \delta_\nu^0 \delta_\alpha^0 \delta_\beta^0 \delta^3,
\label{g-bar-zeta-Gamma2}  
\end{eqnarray}
Eq. (\ref{g-bar-zeta-Gamma}) provides us with
\begin{eqnarray}
[ g_{\mu\nu}, \dot{\bar \zeta}_\rho^\prime ] = - 8 i \frac{1}{(g^{00})^2} \tilde f \phi^{-2}
\delta_\mu^0 \delta_\nu^0 \delta_\rho^0 \bar \zeta^0 \delta^3.
\label{g-bar-zeta-Gamma3}  
\end{eqnarray}
This ETCR enables us to evaluate the ETCR
\begin{eqnarray}
[ \bar \zeta_\mu, \Gamma_{\alpha\beta}^{\gamma\prime} ] = - 4 i \frac{1}{(g^{00})^2} \tilde f \phi^{-2}
g^{0\gamma} \delta_\mu^0 \delta_\alpha^0 \delta_\beta^0 \bar \zeta^0 \delta^3.
\label{g-bar-zeta-Gamma4}  
\end{eqnarray}
Next, let us consider the CCR, $\{ \bar \zeta_\mu, \pi_{\bar \zeta}^{\nu\prime} \}
= i \delta_\mu^\nu \delta^3$, which can be cast to the form
\begin{eqnarray}
\sqrt{-g^\prime} ( g^{\nu\alpha\prime} g^{0\beta\prime} + g^{0\alpha\prime} g^{\nu\beta\prime} ) 
\{ \bar \zeta_\mu, \nabla_\alpha \tilde \zeta_\beta^\prime \}
= - \delta_\mu^\nu \delta^3.
\label{bar-zeta-pi-bar-zeta}  
\end{eqnarray}
This equation, together with Eq. (\ref{g-bar-zeta-Gamma4}), yields the ETCR
\begin{eqnarray}
\{ \bar \zeta_\mu, \dot{\tilde \zeta}_\nu^\prime \} = - \tilde f \Bigl( g_{\mu\nu} 
- \frac{1}{2 g^{00}} \delta_\mu^0 \delta_\nu^0 \Bigr) \delta^3
- 4 i \frac{1}{(g^{00})^2} \tilde f \phi^{-2} \delta_\mu^0 \delta_\nu^0 \bar \zeta^0 \tilde \zeta^0 \delta^3.
\label{bar-zeta-dot-zeta}  
\end{eqnarray}
Then, using Eqs. (\ref{g-bar-zeta-Gamma4}) and (\ref{bar-zeta-dot-zeta}), (\ref{beta-rho-K}) is calculated to
\begin{eqnarray}
[ \beta_\rho, K_{\mu\nu}^\prime ] = i \tilde f \Bigl( \delta_\mu^0 g_{\rho\nu} + \delta_\nu^0 g_{\rho\mu}
- \frac{1}{g^{00}} \delta_\mu^0 \delta_\nu^0 \delta_\rho^0 \Bigr) \delta^3.
\label{beta-rho-K2}  
\end{eqnarray}
With the help of Eqs. (\ref{dot-g-piK}), (\ref{dot-g-0rho}), (\ref{pi-phi-express}) and (\ref{beta-rho-K2}), it turns out that
\begin{eqnarray}
&{}& [ \dot g_{\rho\sigma}, K_{\mu\nu}^\prime ] = i \frac{1}{\gamma} \tilde f \Bigl[ g_{\rho\sigma} g_{\mu\nu}
- 2 g_{\rho(\mu} g_{\nu)\sigma} - \frac{1}{g^{00}} \Bigl( g_{\rho\sigma} \delta_\mu^0 \delta_\nu^0
+ \frac{2}{3} g_{\mu\nu} \delta_\rho^0 \delta_\sigma^0 \Bigr)
\nonumber\\
&{}& + \frac{2}{g^{00}} ( g_{\rho(\mu} \delta_{\nu)}^0 \delta_\sigma^0 
+ g_{\sigma(\mu} \delta_{\nu)}^0 \delta_\rho^0 )
- \frac{4}{3 (g^{00})^2} \delta_\rho^0 \delta_\sigma^0 \delta_\mu^0 \delta_\nu^0 \Bigr] \delta^3
\nonumber\\
&{}& - 8 i \frac{1}{(g^{00})^2} \tilde f \phi^{-2} \delta_\rho^0 \delta_\sigma^0 ( \delta_\mu^0 K^0 \,_\nu
+ \delta_\nu^0 K^0 \,_\mu ) \delta^3.
\label{dot-g-K}  
\end{eqnarray}

To close this section, it is worthwhile to mention that any ETCRs can be in principle calculated by using the CCRs, 
field equations, the BRST transformations and the ETCRs presented thus far although we have not given all the 
ETCRs in this article.

\section{Linearized field equations}

In this section, we analyze asymptotic fields under the assumption that all fields have their own
asymptotic fields and there is no bound state. We also assume that all asymptotic fields are
governed by the quadratic part of the quantum Lagrangian apart from possible renormalization.

We define the gravitational field $\varphi_{\mu\nu}$ on a flat Minkowski metric $\eta_{\mu\nu}$ 
and the scalar fluctuation $\tilde \phi $ on a nonzero fixed scalar field $\phi_0$:
\begin{eqnarray}
g_{\mu\nu} = \eta_{\mu\nu} + \varphi_{\mu\nu},  \qquad
\phi = \phi_0 + \tilde \phi.
\label{Background}  
\end{eqnarray}
For sake of simplicity, we use the same notation for the other asymptotic fields as that for the
interacting fields. Then, up to surface terms the quadratic part of the quantum Lagrangian (\ref{q-Lag}) reads
\begin{eqnarray}
&{}& {\cal L}_q = \frac{1}{12} \phi_0^2 \Bigl( \frac{1}{4} \varphi_{\mu\nu} \Box \varphi^{\mu\nu} 
- \frac{1}{4} \varphi \Box \varphi - \frac{1}{2} \varphi^{\mu\nu} \partial_\mu \partial_\rho \varphi_\nu{}^\rho
+ \frac{1}{2} \varphi^{\mu\nu} \partial_\mu \partial_\nu \varphi \Bigr)
\nonumber\\
&{}& + \frac{1}{6} \phi_0 \tilde \phi \left( - \Box \varphi + \partial_\mu \partial_\nu \varphi^{\mu\nu} \right)
+ \frac{1}{2} \partial_\mu \tilde \phi \partial^\mu \tilde \phi
+ \frac{1}{2} \gamma ( 2 \partial_\mu \partial_\rho \varphi_\nu{}^\rho - \Box \varphi_{\mu\nu}
\nonumber\\
&{}& - \partial_\mu \partial_\nu \varphi ) \bar{K}^{\mu\nu} 
+ \alpha [ ( K_{\mu\nu} - \partial_\mu A_\nu - \partial_\nu A_\mu )^2 - ( K - 2 \partial_\mu A^\mu )^2 ]
\nonumber\\
&{}& - \Bigl( 2 \phi_0 \eta^{\mu\nu} \tilde \phi - \phi_0^2 \varphi^{\mu\nu} + \frac{1}{2} \phi_0^2 \eta^{\mu\nu} \varphi \Bigr) \partial_\mu b_\nu
- i \phi_0^2 \partial_\mu \bar c_\rho \partial^\mu c^\rho 
\nonumber\\
&{}& - b ( K - 2 \partial_\mu A^\mu ) + i \frac{\gamma}{\alpha} \partial_\mu \bar c \partial^\mu c
- \partial_\mu K^{\mu\nu} \beta_\nu + i \partial^\mu \bar \zeta^\nu ( \partial_\mu \tilde \zeta_\nu
+ \partial_\nu \tilde \zeta_\mu ).
\label{Free-Lag}  
\end{eqnarray}
In this and next sections, the spacetime indices $\mu, \nu, \dots$ are raised or lowered by the Minkowski metric $\eta^{\mu\nu}
= \eta_{\mu\nu} = \rm{diag} ( -1, 1, 1, 1)$, and we define $\Box \equiv \eta^{\mu\nu} \partial_\mu \partial_\nu$, $\varphi \equiv \eta^{\mu\nu} 
\varphi_{\mu\nu}$ and $\bar K_{\mu\nu} \equiv K_{\mu\nu} - \frac{1}{2} \eta_{\mu\nu} K$. 

Based on this Lagrangian, it is straightforward to derive the linearized field equations: 
\begin{eqnarray}
&{}& \frac{1}{12} \phi_0^2 \biggl( \frac{1}{2} \Box \varphi_{\mu\nu} - \frac{1}{2} \eta_{\mu\nu} \Box \varphi 
- \partial_\rho \partial_{(\mu} \varphi_{\nu)}{}^\rho + \frac{1}{2} \partial_\mu \partial_\nu \varphi
+ \frac{1}{2} \eta_{\mu\nu} \partial_\rho \partial_\sigma \varphi^{\rho\sigma} \biggr)
\nonumber\\
&{}& + \frac{1}{6} \phi_0 \left( - \eta_{\mu\nu} \Box + \partial_\mu \partial_\nu \right) \tilde \phi
+ \frac{\gamma}{2} \Bigl( 2 \partial_\rho \partial_{(\mu} \bar K_{\nu)}{}^\rho - \Box \bar K_{\mu\nu}
- \eta_{\mu\nu} \partial_\rho \partial_\sigma \bar K^{\rho\sigma} \Bigr)
\nonumber\\
&{}& 
+ \phi_0^2 \Bigl( \partial_{(\mu} b_{\nu)} - \frac{1}{2} \eta_{\mu\nu} \partial_\rho b^\rho \Bigr) = 0.
\label{Linear-Eq1}
\\
&{}& \frac{1}{6} \phi_0 ( \Box \varphi - \partial_\mu \partial_\nu \varphi^{\mu\nu} ) + \Box \tilde \phi 
+ 2 \phi_0 \partial_\mu b^\mu = 0.
\label{Linear-Eq2}
\\
&{}& 2 \partial_\rho \partial_{(\mu} \varphi_{\nu)}{}^\rho - \Box \varphi_{\mu\nu} - \partial_\mu \partial_\nu \varphi 
- \eta_{\mu\nu} ( \partial_\rho \partial_\sigma \varphi^{\rho\sigma} - \Box \varphi ) 
+ \frac{4 \alpha}{\gamma} \Bigl[ K_{\mu\nu} 
\nonumber\\
&{}& - \partial_\mu A_\nu - \partial_\nu A_\mu  - \eta_{\mu\nu} ( K - 2 \partial_\rho A^\rho ) \Bigr]
 + \frac{2}{\gamma} ( - \eta_{\mu\nu} b + \partial_{(\mu} \beta_{\nu)} ) = 0.
\label{Linear-Eq3}
\\
&{}& \partial^\nu \Bigl[ K_{\mu\nu} - \partial_\mu A_\nu - \partial_\nu A_\mu 
 - \eta_{\mu\nu} ( K - 2 \partial_\rho A^\rho ) \Bigr] - \frac{1}{2 \alpha} \partial_\mu b = 0.
\label{Linear-Eq4}
\\
&{}& \partial_\mu \tilde \phi - \frac{1}{2} \phi_0 \Bigl( \partial^\nu \varphi_{\mu\nu} - \frac{1}{2} \partial_\mu \varphi \Bigr) = 0. 
\label{Linear-Eq5}  
\\
&{}& K - 2 \partial_\mu A^\mu = 0. 
\label{Linear-Eq6}  
\\
&{}& \partial_\mu K^{\mu\nu} = 0. 
\label{Linear-Eq7}  
\\
&{}& \Box c^\mu = \Box \bar c_\mu = \Box c = \Box \bar c = 0. 
\label{Linear-Eq8}  
\\
&{}& \Box \tilde \zeta_\mu + \partial_\mu \partial_\nu \tilde \zeta^\nu = \Box \bar \zeta_\mu + \partial_\mu \partial_\nu 
\bar \zeta^\nu = 0. 
\label{Linear-Eq9}  
\end{eqnarray}

Now we are ready to simplify the field equations obtained above. Before doing so, it is more convenient to
make use of the linearized BRST transformations in order to seek for the linearized field equations for
the NL fields $b_\mu, b$ and $\beta_\mu$. Taking the linearized GCT BRST transformation $\delta_B^{(1L)} \bar c_\mu = i b_\mu$
of $\Box \bar c_\mu = 0$ in Eq. (\ref{Linear-Eq8}) gives us 
\begin{eqnarray}
\Box b_\mu = 0,
\label{L-b-rho-eq}  
\end{eqnarray}
which is a linearized analog of Eq. (\ref{b-rho-eq}). Similarly, the linearized WS BRST transformation 
$\delta_B^{(2L)} \bar c = i b$ of $\Box \bar c = 0$ in Eq. (\ref{Linear-Eq8}) produces
\begin{eqnarray}
\Box b = 0.
\label{L-B-eq}  
\end{eqnarray}
Finally, the linearized WS BRST transformation $\delta_B^{(2L)} \bar \zeta_\mu = i \beta_\mu$ of $\Box \bar \zeta_\mu + \partial_\mu \partial_\nu 
\zeta^\nu = 0$ in Eq. (\ref{Linear-Eq9}) yields
\begin{eqnarray}
\Box \beta_\mu + \partial_\mu \partial_\nu \beta^\nu = 0.
\label{L-beta-eq}  
\end{eqnarray}
Of course, Eqs. (\ref{L-b-rho-eq}), (\ref{L-B-eq}) and (\ref{L-beta-eq}) can be also derived by solving the linearized field equations
directly.

Next, operating $\partial^\mu$ on Eq. (\ref{L-beta-eq}) leads to
\begin{eqnarray}
\Box \partial_\mu \beta^\mu = 0.
\label{L-beta-eq2}  
\end{eqnarray}
Moreover, acting $\Box$ on Eq. (\ref{L-beta-eq}) and using Eq. (\ref{L-beta-eq2}), we have
\begin{eqnarray}
\Box^2 \beta_\mu = 0,
\label{L-beta-eq3}  
\end{eqnarray}
which implies that $\beta_\mu$ is a dipole field.
 In a perfectly similar manner, Eq. (\ref{Linear-Eq9}) gives us
\begin{eqnarray}
\Box \partial_\mu \tilde \zeta^\mu = \Box^2 \tilde \zeta_\mu = 0, \quad
\Box \partial_\mu \bar \zeta^\mu = \Box^2 \bar \zeta_\mu = 0.
\label{L-zetas-eq2}  
\end{eqnarray}

Now it is easy to see that with the help of Eqs. (\ref{Linear-Eq6}) and (\ref{Linear-Eq7}), Eq. (\ref{Linear-Eq4}) 
provides\footnote{Note that as a consistency check, the WS BRST transformation of this equation
gives rise to the field equation for $\tilde \zeta_\mu$ in Eq. (\ref{Linear-Eq9}) when we use the
field equation $\Box c = 0$.}  
\begin{eqnarray}
\Box A_\mu + \partial_\mu \partial_\nu A^\nu + \frac{1}{2 \alpha} \partial_\mu b = 0.
\label{A-mu-eq}  
\end{eqnarray}
Given Eq. (\ref{L-B-eq}), this equation shows that the gauge field $A_\mu$ is a dipole field obeying
\begin{eqnarray}
\Box \partial_\mu A^\mu = \Box^2 A_\mu = 0.
\label{Dipol-A-eq}  
\end{eqnarray}
By use of Eq. (\ref{Linear-Eq6}), this equation means that $K$ is a simple field:
\begin{eqnarray}
\Box K = 0.
\label{Simple-K-eq}  
\end{eqnarray}
 
Next, to exhibit that the scalar field $\tilde \phi$ is also a dipole field, let us take the trace of
Eq. (\ref{Linear-Eq3}) whose result can be written as 
\begin{eqnarray}
\Box \varphi - \partial_\mu \partial_\nu \varphi^{\mu\nu} = \frac{1}{\gamma} ( 4 b - \partial_\mu \beta^\mu ),
\label{Box-varphi}  
\end{eqnarray}
where Eq. (\ref{Linear-Eq6}) was utilized. Substituting this equation into Eq. (\ref{Linear-Eq2}) yields 
\begin{eqnarray}
\Box \tilde \phi = - \frac{\phi_0}{6 \gamma} ( 4 b - \partial_\mu \beta^\mu + 12 \gamma \partial_\mu b^\mu ).
\label{Box-tilde-phi}  
\end{eqnarray}
Operating $\Box$ on this equation produces the desired result that $\tilde \phi$ is a dipole field:
\begin{eqnarray}
\Box^2 \tilde \phi = 0,
\label{Dipole-tilde-phi}  
\end{eqnarray}
where we used Eqs.  (\ref{L-b-rho-eq}), (\ref{L-B-eq}) and (\ref{L-beta-eq2}).
The divergence of Eq. (\ref{Linear-Eq5}) takes the form
\begin{eqnarray}
\Box \tilde \phi = \frac{1}{2} \phi_0 ( \partial_\mu \partial_\nu \varphi^{\mu\nu} 
- \frac{1}{2} \Box \varphi ).
\label{Box-tilde-phi2}  
\end{eqnarray}
Using three equations (\ref{Box-varphi}), (\ref{Box-tilde-phi}) and (\ref{Box-tilde-phi2}),
we can describe $\Box \varphi$ and $\partial_\mu \partial_\nu \varphi^{\mu\nu}$ as
\begin{eqnarray}
\Box \varphi &=& \frac{4}{3 \gamma} ( 4 b - \partial_\mu \beta^\mu - 6 \gamma \partial_\mu b^\mu ),
\nonumber\\
\partial_\mu \partial_\nu \varphi^{\mu\nu} &=& \frac{1}{3 \gamma} ( 4 b - \partial_\mu \beta^\mu 
- 24 \gamma \partial_\mu b^\mu ),
\label{Box-tilde-phi3}  
\end{eqnarray}
which imply two equations: 
\begin{eqnarray}
\Box^2 \varphi = \Box \partial_\mu \partial_\nu \varphi^{\mu\nu} = 0.
\label{Box-tilde-phi4}  
\end{eqnarray}

Here it is useful to express $K_{\mu\nu}$ in terms of the other fields by starting with Eq. (\ref{Linear-Eq3}) and utilizing some equations 
obtained thus far, whose result is described as
\begin{eqnarray}
K_{\mu\nu} &=& \partial_\mu A_\nu + \partial_\nu A_\mu + \frac{\gamma}{4 \alpha} \Box \varphi_{\mu\nu} 
- \frac{\gamma}{\alpha \phi_0} \partial_\mu \partial_\nu \tilde \phi 
\nonumber\\
&-& \frac{1}{2 \alpha} \Bigl( \eta_{\mu\nu} b + \partial_{(\mu} \beta_{\nu)} - \frac{1}{2} \eta_{\mu\nu} \partial_\rho \beta^\rho \Bigr). 
\label{K-express}  
\end{eqnarray}

Finally, let us focus on the linearized Einstein equation (\ref{Linear-Eq1}). After some calculations using several equations, it turns out
Eq. (\ref{Linear-Eq1}) can be rewritten into a more compact form:
\begin{eqnarray}
&{}& \Box ( \Box - m^2 ) \varphi_{\mu\nu} + \frac{1}{3 \gamma} m^2 \eta_{\mu\nu} ( 4 b - \partial_\rho \beta^\rho ) 
- \frac{4}{3 \gamma} \partial_\mu \partial_\nu b + \frac{4}{3 \gamma} \partial_\mu \partial_\nu \partial_\rho \beta^\rho
\nonumber\\
&{}& + 8 \partial_\mu \partial_\nu \partial_\rho b^\rho
- 24 m^2 \Bigl( \partial_{(\mu} b_{\nu)} - \frac{1}{6} \eta_{\mu\nu} \partial_\rho b^\rho \Bigr) = 0,    
\label{L-Grav-eq}  
\end{eqnarray}
where we have defined mass squared, $m^2 \equiv \frac{\phi_0^2}{24 \alpha_c} = \frac{\alpha \phi_0^2}{3 \gamma^2}$, 
which demands us to take the positive $\alpha_c$ as assumed before. 
Furthermore, operating $\Box$ on (\ref{L-Grav-eq}), we can obtain the gravitational equation for $\varphi_{\mu\nu}$:
\begin{eqnarray}
\Box^2 \Bigl( \Box - m^2 \Bigr) \varphi_{\mu\nu} = 0.    
\label{L-Grav-eq2}  
\end{eqnarray}

Eq. (\ref{L-Grav-eq2}) implies that there are both massless and massive modes in $\varphi_{\mu\nu}$. In order to disentangle these two modes,
let us act $\Box$ on Eq. (\ref{K-express}): 
\begin{eqnarray}
\Box \Bigl( K_{\mu\nu} - \partial_\mu A_\nu - \partial_\nu A_\mu + \frac{\gamma}{\alpha \phi_0} \partial_\mu \partial_\nu \tilde \phi 
+ \frac{1}{2 \alpha} \partial_{(\mu} \beta_{\nu)} \Bigr) = \frac{\gamma}{4 \alpha} \Box^2 \varphi_{\mu\nu}.    
\label{Box-K-op}  
\end{eqnarray}
This RHS can be further rewritten by using Eqs. (\ref{K-express}) and (\ref{L-Grav-eq}) as
\begin{eqnarray}
&{}& \Box \Bigl( K_{\mu\nu} - \partial_\mu A_\nu - \partial_\nu A_\mu + \frac{\gamma}{\alpha \phi_0} \partial_\mu \partial_\nu \tilde \phi 
+ \frac{1}{2 \alpha} \partial_{(\mu} \beta_{\nu)} \Bigr)
\nonumber\\
&{}&
= m^2 \Bigl[ K_{\mu\nu} - \partial_\mu A_\nu - \partial_\nu A_\mu + \frac{\gamma}{\alpha \phi_0} \partial_\mu \partial_\nu \tilde \phi 
+ \frac{1}{2 \alpha} \partial_{(\mu} \beta_{\nu)} 
+ \frac{1}{6 \alpha} \eta_{\mu\nu} ( b - \partial_\rho \beta^\rho )
\nonumber\\
&{}&
+ \frac{1}{3 \alpha m^2} \partial_\mu \partial_\nu ( b - \partial_\rho \beta^\rho - 6 \gamma \partial_\rho b^\rho ) 
+ \frac{6 \gamma}{\alpha} \Bigl( \partial_{(\mu} b_{\nu)} - \frac{1}{6} \eta_{\mu\nu} \partial_\rho b^\rho \Bigr) \Bigr].
\label{Mass-ghost0}  
\end{eqnarray}
Provided that we take a linear combination of fields given as
\begin{eqnarray}
&{}& \psi_{\mu\nu} = K_{\mu\nu} - \partial_\mu A_\nu - \partial_\nu A_\mu + \frac{\gamma}{\alpha \phi_0} \partial_\mu \partial_\nu \tilde \phi 
+ \frac{1}{2 \alpha} \partial_{(\mu} \beta_{\nu)} 
+ \frac{1}{6 \alpha} \eta_{\mu\nu} ( b - \partial_\rho \beta^\rho )
\nonumber\\
&{}& + \frac{1}{3 \alpha m^2} \partial_\mu \partial_\nu ( b - \partial_\rho \beta^\rho - 6 \gamma \partial_\rho b^\rho ) 
+ \frac{6 \gamma}{\alpha} \Bigl( \partial_{(\mu} b_{\nu)} - \frac{1}{6} \eta_{\mu\nu} \partial_\rho b^\rho \Bigr),
\label{LC-Psi}  
\end{eqnarray}
we find that $\psi_{\mu\nu}$ corresponds to an infamous massive ghost of spin-2 of 5 physical degrees of freedom
since it satisfies the equations of motion
\begin{eqnarray}
(\Box - m^2) \psi_{\mu\nu} =  \psi^\mu_\mu = \partial^\nu \psi_{\mu\nu} = 0.
\label{Eq-mot-Psi}  
\end{eqnarray}

On the other hand, if we choose the following linear combination
\begin{eqnarray}
h_{\mu\nu} &=&  \varphi_{\mu\nu} - \frac{12 \gamma}{\phi_0^2} \psi_{\mu\nu} 
+ \frac{2}{\phi_0} \eta_{\mu\nu} \tilde \phi,
\label{graviton}  
\end{eqnarray}
we find that $h_{\mu\nu}$ obeys the field equation
\begin{eqnarray}
&{}& \Box h_{\mu\nu} = - \frac{4}{3 \gamma m^2} \partial_\mu \partial_\nu ( b - \partial_\rho \beta^\rho 
- 6 \gamma \partial_\rho b^\rho ) - 24 \partial_{(\mu} b_{\nu)},
\nonumber\\
&{}& \partial^\mu h_{\mu\nu} - \frac{1}{2} \partial_\nu h = 0.
\label{graviton-eq}  
\end{eqnarray}
Then, Eq. (\ref{graviton-eq}) implies that $h_{\mu\nu}$ is a dipole field satisfying
\begin{eqnarray}
\Box^2 h_{\mu\nu} = 0.
\label{graviton-eq2}  
\end{eqnarray}
Later we will show that two transverse components of $h_{\mu\nu}$ is nothing but a massless spin-2 graviton.

\section{Analysis of physical states}

Following the standard technique, we can calculate the four-dimensional (anti)commutation 
relations (4D CRs) between asymptotic fields. The point is that the simple pole fields, for instance, 
the Nakanishi-Lautrup field $b_\mu (x)$ obeying $\Box b_\mu = 0$, can be expressed in terms of 
the invariant delta function $D(x)$ as
\begin{eqnarray}
b_\mu (x) = \int d^3 z D(x-z) \overleftrightarrow{\partial}_0^z b_\mu (z).
\label{D-b1}  
\end{eqnarray}
Here the invariant delta function $D(x)$ for massless simple pole fields and its properties
are described as
\begin{eqnarray}
&{}& D(x) = - \frac{i}{(2 \pi)^3} \int d^4 k \, \epsilon (k_0) \delta (k^2) e^{i k x}, \qquad
\Box D(x) = 0,
\nonumber\\
&{}& D(-x) = - D(x), \qquad D(0, \vec{x}) = 0, \qquad 
\partial_0 D(0, \vec{x}) = \delta^3 (x), 
\label{D-function}  
\end{eqnarray}
where $\epsilon (k_0) \equiv \frac{k_0}{|k_0|}$. With these properties, it is easy to see that
the right-hand side (RHS) of Eq. (\ref{D-b1}) is independent of $z^0$, and this fact will be
used in evaluating 4D CRs via the ETCRs shortly.

To illustrate the detail of the calculation, let us evaluate a 4D CR, $[ h_{\mu\nu} (x), b_\rho (y) ]$ explicitly.
Using Eq. (\ref{D-b1}), it can be described as
\begin{eqnarray}
&{}& [ h_{\mu\nu} (x), b_\rho (y) ] 
\nonumber\\
&=& \int d^3 z D(y-z) \overleftrightarrow{\partial}_0^z [ h_{\mu\nu} (x), b_\rho (z) ]
\nonumber\\
&=& \int d^3 z \Bigl( D(y-z) [ h_{\mu\nu} (x), \dot b_\rho (z) ] 
- \partial_0^z D(y-z) [ h_{\mu\nu} (x), b_\rho (z) ] \Bigr).
\label{4D-h&b}  
\end{eqnarray}
As mentioned above, since the RHS of Eq. (\ref{D-b1}) is independent of $z^0$, we put $z^0 = x^0$ in (\ref{4D-h&b}) and use relevant ETCRs
to obtain
\begin{eqnarray}
&{}& [ h_{\mu\nu} (x), b_\rho (z) ] = i \phi_0^{-2} ( \delta_\mu^0 \eta_{\rho\nu}
+ \delta_\nu^0 \eta_{\rho\mu} ) \delta^3 (x-z),
\nonumber\\
&{}& [ h_{\mu\nu} (x), \dot b_\rho (z) ] = - i \phi_0^{-2} ( \delta_\mu^k \eta_{\rho\nu}
+ \delta_\nu^k \eta_{\rho\mu} ) \partial_k \delta^3 (x-z).
\label{4D-h&b2}  
\end{eqnarray}
Substituting Eq. (\ref{4D-h&b2}) into Eq. (\ref{4D-h&b}), we can obtain the 4D CR
\begin{eqnarray}
[ h_{\mu\nu} (x), b_\rho (y) ] 
= i \phi_0^{-2} ( \eta_{\mu\rho} \partial_\nu + \eta_{\nu\rho} \partial_\mu ) D(x-y).
\label{4D-h&b3}  
\end{eqnarray}

In a similar manner, we can calculate the four-dimensional (anti)commutation relations among 
$\psi_{\mu\nu}, h_{\mu\nu}$ and $b_\mu$ etc. To do that, let us note that since $\psi_{\mu\nu}$ obeys a massive 
simple pole equation (\ref{Eq-mot-Psi}), it can be expressed in terms of the invariant delta function 
$\Delta(x; m^2)$ for massive simple pole fields as
\begin{eqnarray}
\psi_{\mu\nu} (x) = \int d^3 z \Delta (x-z; m^2) \overleftrightarrow{\partial}_0^z \psi_{\mu\nu} (z),
\label{psi-Delta}  
\end{eqnarray}
where $\Delta(x; m^2)$ is defined as
\begin{eqnarray}
&{}& \Delta(x; m^2) = - \frac{i}{(2 \pi)^3} \int d^4 k \, \epsilon (k_0) \delta (k^2 + m^2) e^{i k x}, \quad
(\Box - m^2) \Delta(x; m^2) = 0,
\nonumber\\
&{}& \Delta(-x; m^2) = - \Delta(x; m^2), \quad \Delta(0, \vec{x}; m^2) = 0, 
\nonumber\\
&{}& \partial_0 \Delta(0, \vec{x}; m^2) = \delta^3 (x),  \qquad
\Delta(x; 0) = D(x). 
\label{Delta-function}  
\end{eqnarray}

As for $h_{\mu\nu}$, since it is a massless dipole field as can be seen in Eq. (\ref{graviton-eq2}),
it can be described as
\begin{eqnarray}
h_{\mu\nu} (x) = \int d^3 z \left[ D(x-z) \overleftrightarrow{\partial}_0^z h_{\mu\nu} (z)
+ E(x-z) \overleftrightarrow{\partial}_0^z \Box h_{\mu\nu} (z) \right],
\label{E-varphi}  
\end{eqnarray}
where we have introduced the invariant delta function $E(x)$ for massless dipole fields and its properties 
are given by
\begin{eqnarray}
&{}& E(x) = - \frac{i}{(2 \pi)^3} \int d^4 k \, \epsilon (k_0) \delta^\prime (k^2) e^{i k x}, \qquad  
\Box E(x) = D(x),
\nonumber\\
&{}& E(-x) = - E(x), \qquad 
E(0, \vec{x}) = \partial_0 E(0, \vec{x}) = \partial_0^2 E(0, \vec{x}) = 0, 
\nonumber\\ 
&{}& \partial_0^3 E(0, \vec{x}) = - \delta^3 (x). 
\label{E-function}  
\end{eqnarray}
As in Eq. (\ref{D-b1}), we can also show that the RHS of both (\ref{psi-Delta}) and (\ref{E-varphi}) is independent of $z^0$. 

By using the ETCRs summarized in Appendix B, after a lengthy but straightforward calculation, we find the
following 4D CRs among $\psi_{\mu\nu}, h_{\mu\nu}, \tilde \phi, b_\mu, b, \beta_\mu, c^\mu, \bar c_\mu, c$ and $\bar c$:
\begin{eqnarray}
&{}& [ \psi_{\mu\nu} (x), \psi_{\sigma\tau} (y) ] = - i \frac{\phi_0^2}{12 \gamma^2}  \Bigl[ \frac{2}{3} \eta_{\mu\nu} \eta_{\sigma\tau}
- \eta_{\mu\sigma} \eta_{\nu\tau} - \eta_{\mu\tau} \eta_{\nu\sigma}
\nonumber\\
&{}& + \frac{1}{m^2} \bigl( \eta_{\mu\sigma} \partial_\nu \partial_\tau + \eta_{\mu\tau} \partial_\nu \partial_\sigma
+ \eta_{\nu\sigma} \partial_\mu \partial_\tau  + \eta_{\nu\tau} \partial_\mu \partial_\sigma 
\nonumber\\
&{}& - \frac{2}{3} \eta_{\mu\nu} \partial_\sigma \partial_\tau - \frac{2}{3} \eta_{\sigma\tau} \partial_\mu \partial_\nu \bigr)  
- \frac{4}{3 m^4} \partial_\mu \partial_\nu \partial_\sigma \partial_\tau \Bigr] \Delta (x-y; m^2).
\label{4D-CR1}
\\
&{}& [ h_{\mu\nu} (x), h_{\sigma\tau} (y) ] = i \frac{12}{\phi_0^2}  \Bigl[ \eta_{\mu\nu} \eta_{\sigma\tau}
- \eta_{\mu\sigma} \eta_{\nu\tau} - \eta_{\mu\tau} \eta_{\nu\sigma}
\nonumber\\
&{}& + \frac{1}{m^2} \bigl( \eta_{\mu\sigma} \partial_\nu \partial_\tau + \eta_{\mu\tau} \partial_\nu \partial_\sigma
+ \eta_{\nu\sigma} \partial_\mu \partial_\tau  + \eta_{\nu\tau} \partial_\mu \partial_\sigma 
\nonumber\\
&{}& - \frac{2}{3} \eta_{\mu\nu} \partial_\sigma \partial_\tau - \frac{2}{3} \eta_{\sigma\tau} \partial_\mu \partial_\nu \bigr)  
- \frac{4}{3 m^4} \partial_\mu \partial_\nu \partial_\sigma \partial_\tau \Bigr] D (x-y)
\nonumber\\
&{}& + i \frac{12}{\phi_0^2}  \Bigl( \eta_{\mu\sigma} \partial_\nu \partial_\tau + \eta_{\mu\tau} \partial_\nu \partial_\sigma
+ \eta_{\nu\sigma} \partial_\mu \partial_\tau  + \eta_{\nu\tau} \partial_\mu \partial_\sigma 
\nonumber\\
&{}& - \frac{4}{3 m^2} \partial_\mu \partial_\nu \partial_\sigma \partial_\tau \Bigr) E (x-y).
\label{4D-CR2}
\\
&{}& [ h_{\mu\nu} (x), \psi_{\sigma\tau} (y) ] = 0.
\label{4D-CR3}
\\
&{}& [ \psi_{\mu\nu} (x), b_\rho (y) ] = [ \psi_{\mu\nu} (x), b (y) ] = [ \psi_{\mu\nu} (x), \beta_\rho (y) ] = 0.
\label{4D-CR4}
\\
&{}& [ h_{\mu\nu} (x), b_\rho (y) ] = i \phi_0^{-2} ( \eta_{\mu\rho} \partial_\nu + \eta_{\nu\rho} \partial_\mu ) D(x-y).
\label{4D-CR5}
\\
&{}& [ h_{\mu\nu} (x), b (y) ] = [ h_{\mu\nu} (x), \beta_\rho (y) ] = 0.
\label{4D-CR6}
\\
&{}& [ \tilde \phi (x), \tilde \phi (y) ] = - i [ D(x-y) - 2 m^2 E(x-y) ]. 
\label{4D-CR7}
\\
&{}& [ \tilde \phi (x), \psi_{\sigma\tau} (y) ] = 0. 
\label{4D-CR8}
\\
&{}& [ \tilde \phi (x), h_{\sigma\tau} (y) ] =  2i \phi_0^{-1} [ \eta_{\sigma\tau} D(x-y) 
+ 2 \partial_\sigma \partial_\tau E(x-y) ]. 
\label{4D-CR9}
\\
&{}& [ \tilde \phi (x), b (y) ] = - i \frac{\alpha}{\gamma} \phi_0 D(x-y). 
\label{4D-CR10}
\\
&{}& [ \tilde \phi (x), b_\rho (y) ] = [ \tilde \phi (x), \beta_\rho (y) ] = 0. 
\label{4D-CR11}
\\
&{}& \{ c^\mu (x), \bar c_\nu (y) \} = - \phi_0^{-2} \delta_\nu^\mu D(x-y). 
\label{4D-CR12}
\\
&{}& \{ c (x), \bar c (y) \} = \frac{\alpha}{\gamma} D(x-y). 
\label{4D-CR13}  
\end{eqnarray}
In particular, note that the negative sign in front of the RHS of Eq. (\ref{4D-CR1}) implies that the massive spin-2 field
$\psi_{\mu\nu}$ has indefinite norm so it is sometimes called ``{\it{massive ghost}}''.

As usual, the physical Hilbert space $|\rm{phys} \rangle$ is defined by the Kugo-Ojima subsidiary conditions \cite{Kugo-Ojima}
\begin{eqnarray}
\rm{Q_B^{(1)}} |\rm{phys} \rangle = \rm{Q_B^{(2)}} |\rm{phys} \rangle = 0,
\label{Phys-Hilbert}  
\end{eqnarray}
where $\rm{Q_B^{(1)}}$ and $\rm{Q_B^{(2)}}$ are respectively BRST charges corresponding to the GCT and WS BRST
transformations. 

The GCT BRST transformation for the asymptotic fields\footnote{Recall that we use same fields
for the interacting and the asymptotic fields. In this section, all the fields describe the asymptotic ones.}   
is given by
\begin{eqnarray}
&{}& \delta_B^{(1)} \psi_{\mu\nu} = 0,  \qquad
\delta_B^{(1)} h_{\mu\nu} = - ( \partial_\mu c_\nu + \partial_\nu c_\mu ),  \qquad
\delta_B^{(1)} \tilde \phi = 0, 
\nonumber\\
&{}&
\delta_B^{(1)} b_\mu = \delta_B^{(1)} b = \delta_B^{(1)} \beta_\mu = 0,  
\nonumber\\
&{}&
\delta_B^{(1)} \bar c_\mu =  i b_\mu,  \qquad
\delta_B^{(1)} c^\mu = \delta_B^{(1)} c = \delta_B^{(1)} \bar c = 0.
\label{Asym-GCT-BRST}  
\end{eqnarray}
And the WS transformation for the asymptotic fields takes the form
\begin{eqnarray}
&{}& \delta_B^{(2)} \psi_{\mu\nu} = \delta_B^{(2)} h_{\mu\nu} = 0,  \qquad
\delta_B^{(2)} \tilde \phi = - \phi_0 c, 
\nonumber\\
&{}&
\delta_B^{(2)} b_\mu = \delta_B^{(2)} b = \delta_B^{(2)} \beta_\mu = 0,  
\nonumber\\
&{}&
\delta_B^{(2)} \bar c =  i b,  \qquad
\delta_B^{(2)} c^\mu = \delta_B^{(2)} \bar c_\mu = \delta_B^{(2)} c = 0.
\label{Asym-WS-BRST}  
\end{eqnarray}

Given the physical state conditions (\ref{Phys-Hilbert}) and the two BRST transformations (\ref{Asym-GCT-BRST}) and 
(\ref{Asym-WS-BRST}), it is easy to clarify the physical content of the theory under consideration: The physical modes
are composed of both a spin-2 massive ghost $\psi_{\mu\nu}$ of mass $m$ which has five physical degrees of freedom,
and a spin-2 massless graviton which corresponds to two components of $h_{\mu\nu}$ (for instance, in the specific
Lorentz frame $p_\mu = ( p, 0, 0, p)$, the graviton corresponds to $\frac{1}{\sqrt{2}} ( h_{11} - h_{22} )$ and $h_{12}$.). 
On the other hand, the remaining four components of $h_{\mu\nu}$, $b_\mu$, $c^\mu$ and $\bar c_\mu$ belong to 
a GCT-BRST quartet while $\tilde \phi$, $b$, $c$ and $\bar c$ does a WS-BRST quartet. These quartets
appear in the physical subspace only as zero norm states by the Kugo-Ojima subsidiary conditions (\ref{Phys-Hilbert}).
It is worthwhile to stress that the massive ghost with indefinite norm appears in the physical Hilbert space so
the unitarity of the physical S-matrix is explicitly violated in the present theory.

\section{Conclusion}

In this article, on the basis of the BRST formalism we have presented the manifestly covariant canonical operator formalism 
of a Weyl invariant gravity where the classical Lagrangian is constituted of the well-known conformal gravity and a Weyl invariant 
scalar-tensor gravity. Once the unitary gauge, $\phi = \sqrt{\frac{3}{4 \pi G}}$ for the Weyl symmetry is taken,
the classical theory becomes equivalent to general relativity plus conformal gravity, so at low energies our theory properly reduces
to Einstein's general relativity while at high eneries it reduces to conformal gravity where a local scale symmetry, or
equivalently the Weyl symmetry, emerges in addition to the general coordinate invariance.  This fact would
give us some distict phenomenological consequences from those obtained through only Einstein's general relativity 
for inflation and the scale invariant spectrum of the Cosmic Microwave Background (CMB) radiation etc.  

One of the important ingredients in the present formalism lies in the choice of gauge conditions for three local symmetries,
those are, the general coordinate invariance, the Weyl symmetry and the St\"{u}ckelberg symmetry.
We have required that the proper gauge conditions should not only fix the gauge symmetries completely but also
give us the maximal global symmetry. As a result, we are led to selecting the extended de Donder gauge condition,
the traceless gauge condition and the K-gauge condition. We think that these gauge conditions are almost unique
up to terms involving the NL fields multiplied by the gauge parameters. 

A question often asked in gravitational theories is that global symmetries such as the Poincar${\rm{\acute{e}}}$-like 
$\IOSp(8|8)$ symmetry are effective symmetries existing only at low energies or exact symmetries holding
even at high energies. To address this question, for instance, one has to construct a renormalizable quantum gravity 
and show that such global symmetires still exist in such a ultraviolet (UV) complete quantum gravity. Since quantum
conformal gravity under consideration is a renormalizable theory as long as the Weyl symmetry is free from Weyl anomaly, 
the Poincar${\rm{\acute{e}}}$-like $\IOSp(8|8)$ symmetry is not an effective but an exact global symmetry. Moreover, this symmetry is 
closely related to purely quantum fields such as ghosts and the Nakanishi-Lautrup fields, so it is not violated by black hole's no-hair 
theorem \cite{MTW}.   

As future's works, we wish to comment on two important issues. One of them is of course related to the issue of the massive
ghost which violates the unitarity of the quantum theory. Recently it has been clearly shown that the Lee-Wick's prescription 
\cite{LW1, LW2} dealing with the ghost fields does not work well at least within the standard framework of quantum field theories \cite{Kugo-Kubo}.  
Thus, if our theory makes sense as a quantum field theory, we cannot rely on the Lee-Wick's prescription any longer and
should develop a new dynamical mechanism. Regarding this problem, it might be useful to recall that as mentioned in Section 3
the quantum conformal gravity is in a sense similar to the QCD while the quantum Einstein's gravity is similar to the QED.
It is known that in the QCD, gluons and quarks are confined to the unphysical sector.  
Thus, we could conjecture that the global symmeties existing in the quantum conformal gravity might play a role to make 
the massive ghost be confined to the unphysical sector. 

The other important issue is relevant to Weyl anomaly. More recently, this issue has been considered in Ref. \cite{W-anomaly}
where it is mentioned that there is no Weyl anomaly in the unbroken phase ($\langle \phi \rangle= 0$) but Weyl anomaly appears in the broken phase 
($\langle \phi \rangle \neq 0$) in Weyl geometry which is a generalization of Riemann geometry. We wish to understand whether 
the similar results hold even in our theory formulated in Riemann geometry. Actually, the fact that the Weyl symmetry could be maintained and manifest 
even at the quantum level in Riemann geometry has been already discussed in Refs. \cite{Englert, Shaposhnikov1, Shaposhnikov2, Codello}.
We would like to return these two issues in future.

\section*{Acknowledgment}

This work is supported in part by the JSPS Kakenhi Grant No. 21K03539.

\appendix
\addcontentsline{toc}{section}{Appendix~\ref{app:scripts}: Training Scripts}
\section*{Appendix}
\label{app:scripts}
\renewcommand{\theequation}{A.\arabic{equation}}
\setcounter{equation}{0}

\section{Derivation of Eq. (\ref{dot-Amu-b})}
\def\T{\text{T}}

In this appendix, we present a derivation of Eq. (\ref{dot-Amu-b}).  

First, we make use of the translational invariance of the theory under consideration. The translational invariance 
requires the validity of the following equation for a generic field $\Phi (x)$:   
\begin{eqnarray}
[ \Phi (x), P_\rho ] = i \partial_\rho \Phi (x),  
\label{Transl-Phi}  
\end{eqnarray}
where $P_\rho$ is the generator of the translation defined as
\begin{eqnarray}
P_\rho = \int d^3 x \tilde g^{0\lambda} \phi^2 \partial_\lambda b_\rho.  
\label{Transl-Gen}  
\end{eqnarray}

Next, taking the specific case $\Phi (x) = A_\mu (x)$, we have
\begin{eqnarray}
[ A_\mu (x), P_\rho ] = [ A_\mu (x), \int d^3 x^\prime \tilde g^{0\lambda\prime} \phi^{2\prime}
\partial_\lambda b_\rho^\prime ] = i \partial_\rho A_\mu (x).
\label{Transl-A}  
\end{eqnarray}
Then, putting $x^0 = x^{0\prime}$ and using $[ A_\mu (x), \tilde g^{0\lambda\prime} \phi^{2\prime}]
=0$ produces 
\begin{eqnarray}
\int d^3 x^\prime \tilde g^{0\lambda\prime} \phi^{2\prime} [ A_\mu, \partial_\lambda b_\rho^\prime ] 
= i \partial_\rho A_\mu (x).
\label{Transl-A2}  
\end{eqnarray}
By means of the extended de Donder condition (\ref{Ext-de-Donder}) and Eq. (\ref{Var-b-rho}), Eq. (\ref{Transl-A2}) can be rewritten as 
\begin{eqnarray}
\int d^3 x^\prime \tilde g^{00\prime} \phi^{2\prime} [ A_\mu, \dot b_\rho^\prime ] 
= i [ \partial_\rho A_\mu + \partial_0 ( \tilde g^{00} \phi^2 ) \tilde f \phi^{-2} \delta_\mu^0 A_\rho ].
\label{Transl-A3}  
\end{eqnarray}
This equation can be easily solved to be
\begin{eqnarray}
[ A_\mu, \dot b_\rho^\prime ] 
= i \tilde f \phi^{-2} [ \partial_\rho A_\mu + \partial_0 ( \tilde g^{00} \phi^2 ) \tilde f \phi^{-2} \delta_\mu^0 A_\rho ] \delta^3 
+ F_{\mu\rho}^k \partial_k ( \tilde f \phi^{-2} \delta^3 ),
\label{Transl-sol}  
\end{eqnarray}
where $F_{\mu\rho}^k$ is an arbitrary function.

To fix the function $F_{\mu\rho}^k$, let us impose the consistency condition 
\begin{eqnarray}
[ g^{\mu\nu} \partial_\mu A_\nu, b_\rho^\prime ] = 0.
\label{Consist-A}  
\end{eqnarray}
This consistency condition comes from the following argument: First, note that since $\nabla_\mu ( g^{\mu\nu} \phi^2 A_\nu )$ is a scalar,
we have
\begin{eqnarray}
[ \nabla_\mu ( g^{\mu\nu} \phi^2 A_\nu ), b_\rho^\prime ] = 0.
\label{Consist-A-1}  
\end{eqnarray}
Then, the extended de Donder condition (\ref{Ext-de-Donder}) allows us to rewrite $\nabla_\mu ( g^{\mu\nu} \phi^2 A_\nu )$ as
\begin{eqnarray}
\nabla_\mu ( g^{\mu\nu} \phi^2 A_\nu ) = \frac{1}{\sqrt{-g}} \partial_\mu ( \tilde g^{\mu\nu} \phi^2 A_\nu )
=  g^{\mu\nu} \phi^2 \partial_\mu A_\nu.
\label{Consist-A-2}  
\end{eqnarray}
Thus, together with $[ \phi, b_\rho^\prime ] = 0$ in Eq. (\ref{phi&b}), we find that Eq. (\ref{Consist-A-1}) provides Eq. (\ref{Consist-A}). 

After some calculations, Eq. (\ref{Consist-A}) turns out to lead to an equation for the arbitrary function $F_{\mu\rho}^k$:
\begin{eqnarray}
g^{0\mu} F_{\mu\rho}^k = - i g^{0k} A_\rho, 
\label{Consist-A2}  
\end{eqnarray}
which has the unique solution given by 
\begin{eqnarray}
F_{\mu\rho}^k = - i ( 2 \tilde f \tilde g^{0k} \delta_\mu^0 - \delta_\mu^k ) A_\rho.
\label{Consist-A-sol}  
\end{eqnarray}
Substituting Eq. (\ref{Consist-A-sol}) into Eq. (\ref{Transl-sol}), we can obtain
\begin{eqnarray}
[ A_\mu, \dot b_\rho^\prime ] &=& - i ( 2 \tilde f \tilde g^{0i} \delta_\mu^0 - \delta_\mu^i ) 
A_\rho \partial_i ( \tilde f \phi^{-2} \delta^3 ) 
+ i \tilde f \phi^{-2} [ \partial_\rho A_\mu 
\nonumber\\
&+& \partial_0 ( \tilde g^{00} \phi^2 ) \tilde f \phi^{-2} \delta_\mu^0 \partial_0 A_\rho ] \delta^3.  
\label{dot-Amu-b0}  
\end{eqnarray}
Finally, using Eqs. (\ref{dot-Amu-b0}) and (\ref{Var-b-rho}), we arrive at the desired equation
(\ref{dot-Amu-b}).

\renewcommand{\theequation}{B.\arabic{equation}}
\setcounter{equation}{0}

\section{Various equal-time commutation relations in the linearized level}

In this Appendix, we simply write down various equal-time (anti)commutation relations (ETCRs)
which are useful in deriving the four-dimensional commutation relations (4D CRs)
in Eqs. (\ref{4D-CR1})-(\ref{4D-CR13}). These ETCRs can be derived by using the canonical
(anti)commutation relations (CRs), the BRST transformations and the linearized field equations.
(The details of the derivation are omitted in this article.)

\begin{eqnarray}
&{}& [ \dot \varphi_{\mu\nu}, \varphi_{\sigma\tau}^\prime ] = 16 i \phi_0^{-2} \delta_\mu^0 \delta_\nu^0 
\delta_\sigma^0 \delta_\tau^0 \delta^3,
\nonumber\\
&{}& [ \varphi_{\mu\nu}, \dot{\tilde \phi}^\prime ] = 4 i \phi_0^{-1} \delta_\mu^0 \delta_\nu^0 \delta^3, \qquad
[ \varphi_{\mu\nu}, \ddot{\tilde \phi}^\prime ] = - 4 i \phi_0^{-1} ( \delta_\mu^0 \delta_\nu^i
+ \delta_\mu^i \delta_\nu^0 ) \partial_i \delta^3,
\nonumber\\
&{}& [ \varphi_{\mu\nu}, \dddot{\tilde \phi}^\prime ] = 4 i \phi_0^{-1} ( - m^2 \eta_{\mu\nu} 
+ 2 \delta_\mu^0 \delta_\nu^0 \Delta + \delta_\mu^i \delta_\nu^j \partial_i \partial_j ) \delta^3,
\nonumber\\
&{}& [ \varphi_{\mu\nu}, \dot A_\sigma^\prime ] = 0, \qquad
[ \varphi_{\mu\nu}, \ddot A_\sigma^\prime ] = - i \frac{1}{2 \gamma} \eta_{\mu\nu} \delta_\sigma^0 \delta^3,
\nonumber\\
&{}& [ \dot \varphi_{\mu\nu}, K_{\sigma\tau}^\prime ] = - i \frac{1}{\gamma}  \Bigl[ \eta_{\mu\nu} \eta_{\sigma\tau}
- \eta_{\mu\sigma} \eta_{\nu\tau} - \eta_{\mu\tau} \eta_{\nu\sigma} + \eta_{\mu\nu} \delta_\sigma^0 \delta_\tau^0
+ \frac{2}{3} \eta_{\sigma\tau} \delta_\mu^0 \delta_\nu^0 
\nonumber\\
&{}& - ( \eta_{\mu\sigma} \delta_\tau^0 + \eta_{\mu\tau} \delta_\sigma^0 ) \delta_\nu^0
- ( \eta_{\nu\sigma} \delta_\tau^0 + \eta_{\nu\tau} \delta_\sigma^0 ) \delta_\mu^0
- \frac{4}{3} \delta_\mu^0 \delta_\nu^0 \delta_\sigma^0 \delta_\tau^0 \Bigr] \delta^3,
\nonumber\\
&{}& [ \varphi_{\mu\nu}, b_\rho^\prime ] = i \phi_0^{-2} ( \delta_\mu^0 \eta_{\rho\nu} + \delta_\nu^0 \eta_{\rho\mu} ) \delta^3,
\qquad
[ \varphi_{\mu\nu}, \dot b_\rho^\prime ] = - i \phi_0^{-2} ( \delta_\mu^i \eta_{\rho\nu} + \delta_\nu^i \eta_{\rho\mu} ) \partial_i \delta^3,
\nonumber\\
&{}& [ \varphi_{\mu\nu}, b^\prime ] = 0,  \qquad
[ \varphi_{\mu\nu}, \dot b^\prime ] = - 2 i \frac{\alpha}{\gamma} \eta_{\mu\nu} \delta^3.
\label{M-ET1}
\end{eqnarray}
\begin{eqnarray}
&{}& [ \dot{\tilde \phi}, \tilde \phi^\prime ] = i \delta^3, \qquad
[ \dot{\tilde \phi}, \ddot{\tilde \phi}^\prime ] = i ( \Delta - 2 m^2 ) \delta^3, \qquad
[ \ddot{\tilde \phi}, \tilde \phi^\prime ] = [ \ddot{\tilde \phi}, \ddot{\tilde \phi}^\prime ] = 0, 
\nonumber\\
&{}& [ \dddot{\tilde \phi}, \ddot{\tilde \phi}^\prime ] = i \Delta ( \Delta - 4 m^2 ) \delta^3,
\nonumber\\
&{}& [ \dot{\tilde \phi}, K_{\sigma\tau}^\prime ] = i \frac{\phi_0}{6 \gamma} ( \eta_{\sigma\tau} + \delta_\sigma^0
\delta_\tau^0 ) \delta^3,  \qquad
[ \ddot{\tilde \phi}, K_{\sigma\tau}^\prime ] = i \frac{\phi_0}{6 \gamma} ( \delta_\sigma^0 \delta_\tau^i
+ \delta_\tau^0 \delta_\sigma^i ) \partial_i \delta^3,
\nonumber\\
&{}& 
[ \dddot{\tilde \phi}, K_{\sigma\tau}^\prime ] = i \frac{\phi_0}{6 \gamma} [ ( \eta_{\sigma\tau}  
+ 2 \delta_\sigma^0 \delta_\tau^0 ) \Delta + \delta_\sigma^i \delta_\tau^j \partial_i \partial_j ] \delta^3, 
\nonumber\\
&{}& [ \dot{\tilde \phi}, A_\sigma^\prime ] = 0,  \qquad
[ \ddot{\tilde \phi}, A_\sigma^\prime ] = [ \tilde \phi, \ddot{A_\sigma}^\prime ] 
= i \frac{\phi_0}{4 \gamma} \delta_\sigma^0 \delta^3, 
\nonumber\\
&{}& 
[ \dddot{\tilde \phi}, A_\sigma^\prime ] = - i \frac{\phi_0}{4 \gamma} \delta_\sigma^i \partial_i \delta^3, \qquad
[ \ddot{\tilde \phi}, \ddot A_\sigma^\prime ] = i \frac{\phi_0}{2 \gamma} \delta_\sigma^0 \Delta \delta^3, 
\nonumber\\
&{}& [ \dot{\tilde \phi}, b^\prime ] = - i \frac{\alpha}{\gamma} \phi_0 \delta^3,  \qquad
[ \ddot{\tilde \phi}, \dot b^\prime ] = i \frac{\alpha}{\gamma} \phi_0 \Delta \delta^3.
\label{M-ET2}
\end{eqnarray}
\begin{eqnarray}
&{}& [ \dot K_{\mu\nu}, K_{\sigma\tau}^\prime ] = i \frac{\phi_0^2}{12 \gamma^2}  \Bigl[ - \frac{2}{3} \eta_{\mu\nu} \eta_{\sigma\tau}
+ \eta_{\mu\sigma} \eta_{\nu\tau} + \eta_{\mu\tau} \eta_{\nu\sigma}
- \frac{2}{3} ( \delta_\mu^0 \delta_\nu^0 \eta_{\sigma\tau} + \delta_\sigma^0 \delta_\tau^0 \eta_{\mu\nu} )
\nonumber\\
&{}& + \delta_\mu^0 \delta_\sigma^0 \eta_{\nu\tau} + \delta_\mu^0 \delta_\tau^0 \eta_{\nu\sigma}
+ \delta_\nu^0 \delta_\sigma^0 \eta_{\mu\tau} + \delta_\nu^0 \delta_\tau^0 \eta_{\mu\sigma}
+ \frac{4}{3} \delta_\mu^0 \delta_\nu^0 \delta_\sigma^0 \delta_\tau^0 \Bigr] \delta^3,
\nonumber\\
&{}& [ K_{\mu\nu}, \dot A_\sigma^\prime ] = [ K_{\mu\nu}, \ddot A_\sigma^\prime ] = 0,
\nonumber\\
&{}& [ K_{\mu\nu}, \beta_\sigma^\prime ] = i ( \delta_\mu^0 \eta_{\sigma\nu} + \delta_\nu^0 \eta_{\sigma\mu}
+ \delta_\mu^0 \delta_\nu^0 \delta_\sigma^0 ) \delta^3,
\nonumber\\
&{}& [ K_{\mu\nu}, \dot \beta_\sigma^\prime ] = i ( \delta_\mu^0 \delta_\nu^0 \delta_\sigma^i
+ \delta_\mu^0 \delta_\nu^i \delta_\sigma^0 + \delta_\mu^i \delta_\nu^0 \delta_\sigma^0 
+ \eta_{\mu\sigma} \delta_\nu^i + \eta_{\nu\sigma} \delta_\mu^i ) \partial_i \delta^3,
\nonumber\\
&{}& [ K_{\mu\nu}, \dot \beta_\sigma^\prime ] = i ( \delta_\mu^0 \delta_\nu^0 \delta_\sigma^i
+ \delta_\mu^0 \delta_\nu^i \delta_\sigma^0 + \delta_\mu^i \delta_\nu^0 \delta_\sigma^0 
+ \eta_{\mu\sigma} \delta_\nu^i + \eta_{\nu\sigma} \delta_\mu^i ) \partial_i \delta^3,
\nonumber\\
&{}& [ K_{\mu\nu}, \ddot \beta_\sigma^\prime ] = i [ ( \delta_\mu^0 \eta_{\nu\sigma}
+ \delta_\nu^0 \eta_{\mu\sigma} + 2 \delta_\mu^0 \delta_\nu^0 \delta_\sigma^0 ) \Delta
+ ( \delta_\mu^0 \delta_\nu^i \delta_\sigma^j + \delta_\mu^i \delta_\nu^0 \delta_\sigma^j
\nonumber\\
&{}& + \delta_\mu^i \delta_\nu^j \delta_\sigma^0 ) \partial_i \partial_j ] \delta^3,
\nonumber\\
&{}& [ K_{\mu\nu}, \dddot{\beta}_\sigma^\prime ] = i \{ [ \delta_\mu^i \eta_{\nu\sigma}
+ \delta_\nu^i \eta_{\mu\sigma} + 2 ( \delta_\mu^0 \delta_\nu^0 \delta_\sigma^i 
+ \delta_\mu^0 \delta_\nu^i \delta_\sigma^0 + \delta_\mu^i \delta_\nu^0 \delta_\sigma^0 ) ] \Delta 
\nonumber\\
&{}& + \delta_\mu^i \delta_\nu^j \delta_\sigma^k 
\partial_j \partial_k \} \partial_i \delta^3.
\label{M-ET3}
\end{eqnarray}

\begin{eqnarray}
&{}& [ \dot A_\mu, A_\sigma^\prime ] = i \frac{1}{4 \alpha} ( \eta_{\mu\sigma} + \delta_\mu^0 \delta_\sigma^0 )
\delta^3,  \qquad
[ \ddot A_\mu, A_\sigma^\prime ] = i \frac{1}{4 \alpha} ( \delta_\mu^0 \delta_\sigma^i 
+ \delta_\mu^i \delta_\sigma^0 ) \partial_i \delta^3,
\nonumber\\
&{}& [ \ddot A_\mu, \dot A_\sigma^\prime ] = - i \frac{1}{4 \alpha} [ ( \eta_{\mu\sigma} + 2 \delta_\mu^0 \delta_\sigma^0 )
\Delta + \delta_\mu^i \delta_\sigma^j \partial_i \partial_j ] \delta^3,
\nonumber\\
&{}& [ A_\mu, b^\prime ] = - i \frac{1}{2} \delta_\mu^0 \delta^3,  \qquad
[ \dot A_\mu, b^\prime ] = - i \frac{1}{2} \delta_\mu^i \partial_i \delta^3,  \qquad
[ \dot A_\mu, \dot b^\prime ] = i \frac{1}{2} \delta_\mu^0 \Delta \delta^3,
\nonumber\\
&{}& [ A_\mu, \dot \beta_\sigma^\prime ] = - i \Bigl( \eta_{\mu\sigma} + \frac{1}{2} 
\delta_\mu^0 \delta_\sigma^0 \Bigr) \delta^3,  \qquad
[ A_\mu, \ddot \beta_\sigma^\prime ] =  i \frac{1}{2} ( \delta_\mu^0 \delta_\sigma^i
+ \delta_\mu^i \delta_\sigma^0 ) \partial_i \delta^3,
\nonumber\\
&{}& [ A_\mu, \dddot \beta_\sigma^\prime ] = - i \Bigl[ ( \eta_{\mu\sigma} + \delta_\mu^0 \delta_\sigma^0 ) \Delta
+ \frac{1}{2} \delta_\mu^i \delta_\sigma^j \partial_i \partial_j \Bigr] \delta^3,
\nonumber\\
&{}& [ \dot A_\mu, \dddot \beta_\sigma^\prime ] = - i ( \delta_\mu^0 \delta_\sigma^i 
+ \delta_\mu^i \delta_\sigma^0 ) \partial_i \Delta \delta^3.
\label{M-ET4}
\end{eqnarray}

\begin{eqnarray}
&{}& \{ \dot{\bar \zeta}_\mu, \tilde \zeta_\sigma^\prime \} = - \Bigl( \eta_{\mu\sigma} 
+ \frac{1}{2} \delta_\mu^0 \delta_\sigma^0 \Bigr) \delta^3,  \qquad
\{ \ddot{\bar \zeta}_\mu, \tilde \zeta_\sigma^\prime \} = - \frac{1}{2} ( \delta_\mu^0 \delta_\sigma^i
+  \delta_\mu^i \delta_\sigma^0 ) \partial_i \delta^3,
\nonumber\\
&{}& \{ c^\mu, \dot{\bar c}_\sigma^\prime \} = \phi_0^{-2} \delta_\sigma^\mu \delta^3,  \qquad
\{ c, \dot{\bar c}^\prime \} = - \frac{\alpha}{\gamma} \delta^3.
\label{M-ET5}
\end{eqnarray}


\end{document}